\documentclass[conference]{IEEEtran}
\IEEEoverridecommandlockouts
\usepackage{cite}
\usepackage{caption}
\usepackage{amsmath,amssymb,amsfonts}
\usepackage{algorithm}
\usepackage{algorithmic}
\usepackage{graphicx}
\usepackage{textcomp}
\usepackage{xcolor}
\usepackage{booktabs}
\usepackage{hyperref}
\def\BibTeX{{\rm B\kern-.05em{\sc i\kern-.025em b}\kern-.08em
    T\kern-.1667em\lower.7ex\hbox{E}\kern-.125emX}}
\begin{document}

\title{Who Restores the Peg? A Mean-Field Game Approach to Model Stablecoin Market Dynamics}

\author{\IEEEauthorblockN{Hardhik Mohanty}
\IEEEauthorblockA{
Viterbi School of Engineering\\
\textit{University of Southern California}\\
Los Angeles, California \\
hmohanty@usc.edu}
\and
\IEEEauthorblockN{Bhaskar Krishnamachari}
\IEEEauthorblockA{
Viterbi School of Engineering\\
\textit{University of Southern California}\\
Los Angeles, California \\
bkrishna@usc.edu}
}

\maketitle

\begin{abstract}
USDC and USDT are the dominant stablecoins pegged to \$1 with a total market capitalization of over \$300B and rising. Stablecoins make dollar value globally accessible with secure transfer and settlement. Yet in practice, these stablecoins experience periods of stress and de-pegging from their \$1 target, posing significant systemic risks. The behavior of market participants during these stress events and the collective actions that either restore or break the peg are not well understood.
This paper addresses the question: \textit{``who restores the peg?''}. We develop a dynamic, agent-based mean-field game framework for fiat-collateralized stablecoins, in which a large population of arbitrageurs and retail traders strategically interact across primary and secondary markets during a de-peg episode. The key advantage of this equilibrium formulation is that it endogenously maps market frictions into a market-clearing price path and implied net order flows, allowing us to attribute peg-reverting pressure by channel and to stress-test when a given infrastructure becomes insufficient for recovery.
Using three historical de-peg events, we show that the calibrated equilibrium reproduces observed recovery half-lives and yields an order flow decomposition in which system-wide stress is predominantly stabilized by primary-market arbitrage. Finally, a quantitative sensitivity analysis identifies a non-linear breakdown threshold, beyond which a de-peg becomes markedly slower to reverse.
\end{abstract}


\section{Introduction}\label{sec:introduction}
Fiat-collateralized stablecoins such as USDC and USDT have become a fundamental component of the digital asset ecosystem. With the combined market capitalization exceeding \$300 billion as of 2025, these assets serve as a critical bridge between the traditional financial system and the decentralized finance (DeFi) sector \cite{gorton2023taming}. They function as the primary unit of account, a medium of exchange for 24/7 global settlement, and a core form of collateral for decentralized lending and derivatives protocols. The perceived stability of stablecoins supports the valuation and liquidity of thousands of other digital assets. However, this reliance introduces significant risk as these stablecoins are not immune to periods of market stress and can ``de-peg" from their \$1 target, as evidenced by the acute de-pegging of USDC in March 2023 \cite{abraham2024crypto}. The mechanism and underlying infrastructure that control these de-peg and re-peg dynamics, particularly the collective behavior of market participants, are not well understood. This paper attempts to address the critical question of who restores the peg?

\begin{figure}[!t]
    \centering
    \includegraphics[width=0.9\columnwidth]{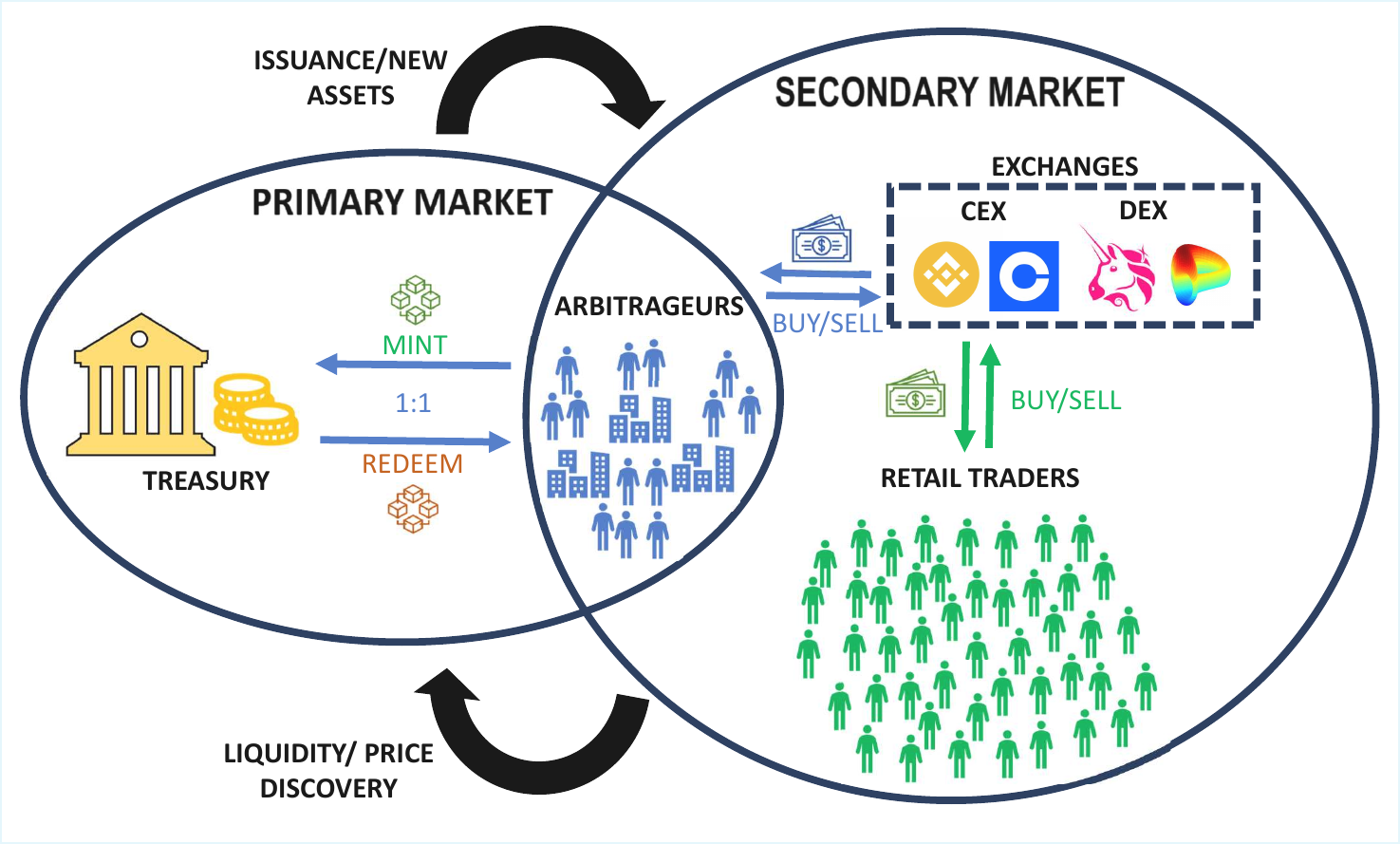}
    \caption{Market structure for fiat-collateralized stablecoins. The primary market involves the treasury and arbitrageurs for 1:1 minting and redeeming, while the secondary market includes exchanges and retail traders for buying and selling assets.}
    \label{fig:market_structure}
\end{figure}

The stability of fiat-collateralized stablecoins is maintained by two independent mechanisms as demonstrated in Figure~\ref{fig:market_structure}: a primary market where authorized participants (arbitrageurs) can mint or redeem the stablecoin directly with the treasury at the \$1 peg, and a diverse ecosystem of secondary markets (e.g., centralized and decentralized exchanges) where the asset is freely traded \cite{lyons2023keeps}. The effectiveness of this pegging mechanism is critically dependent on the design of the primary arbitrage channel characterized by accessibility, cost, and processing speed \cite{potter2024drives}. These markets are populated by a heterogeneous collection of multiple agents, ranging from arbitrageurs to retail traders. 
While arbitrage is a well-known economic mechanism for maintaining the \$1 peg, the tipping point at which this mechanism fails due to congestion or reduced market liquidity remains unknown. It remains unclear how collective actions of market participants aggregate to create a specific breakdown point at which the peg collapses, despite rational arbitrage incentives. Our model attempts to quantify this non-linear threshold.

Rigorously modeling this system is challenging because it is a high-dimensional multi-population game with rational agents interacting dynamically in a non-stationary environment \cite{magninosolving}. Standard multi-agent reinforcement learning (MARL) or agent-based models (ABM) become computationally intractable as the number of agents ($N$) grows, making it difficult to solve for true Nash equilibria. The Mean-Field Game (MFG) approach overcomes this by approximating the population as a continuum ($N \to \infty$), reducing the complexity to a tractable fixed-point problem \cite{cacace2021policy}. 
We adopt an MFG formulation to provide a dynamic, agent-based representation of stablecoin de-peg episodes. This approach offers critical insights that are inaccessible through existing methodologies \cite{lauriere2022learning, bensoussan2016linear}. 

While prior works such as \cite{lyons2023keeps} and \cite{daud2024primary} infer reversion dynamics from empirical data, our framework derives these market flows endogenously by optimizing the strategic behavior of separate, homogeneous populations of agents interacting across primary and secondary markets. This endogenous derivation allows us to identify the specific non-linear thresholds at which primary market frictions overwhelm arbitrage capacity, a transition that reduced-form models cannot capture. Unlike the static equilibrium analysis in \cite{potter2024drives}, our dynamic formulation characterizes the entire temporal profile of peg restoration, including the calculation of recovery half-lives. Furthermore, in contrast to heuristic simulators like DAISIM~\cite{bhat2021daisim} and \cite{calandra2025algorithmic} which rely on pre-defined behavior rules, our model computes a formal equilibrium with quantified exploitability. Consequently, we can evaluate peg stability through the lens of $\varepsilon$-Nash robustness, providing a rigorous metric for how multiple rational, homogeneous population of agents respond to systemic stress across stablecoin markets.

The contributions of this paper are threefold:
\begin{itemize}
    \item First, we contribute to the literature on digital asset stability by developing a dynamic mean-field game framework that serves as a tractable agent-based model for stablecoin de-peg episodes. To our knowledge, this is the first study to analyze these events through an equilibrium-based representation of dynamic strategic interactions between arbitrageurs and retail traders across both primary and secondary markets.
    \item Second, we provide a rigorous assessment of the model by calibrating it to three major historical events, including the USDT and USDC de-pegs of 2022 and 2023. We demonstrate that the framework accurately replicates observed price trajectories and recovery half-lives in ways that standard reduced-form regressions fail to achieve.
    \item Finally, our findings should be of interest to stablecoin issuers and regulators as we identify a non-linear stability threshold for primary-market access. Our analysis reveals that peg restoration is highly sensitive to specific frictions in primary redemption rails and that secondary market liquidity acts only as a secondary amplifier once these primary bottlenecks are crossed.
\end{itemize}

The rest of the paper is structured as follows: Section~\ref{sec:related_work} reviews the related literature. Section~\ref{sec:methodology} presents the dynamic, agent-based mean-field game framework and the market dynamics. Section~\ref{sec:simulation_parameters} describes the baseline calibration and data extraction. Section~\ref{sec:experimental_results} reports the experimental results. Section~\ref{sec:conclusion} concludes and outlines directions for future work.

\begin{figure}[!t]
    \centering
    \includegraphics[width=0.95\columnwidth]{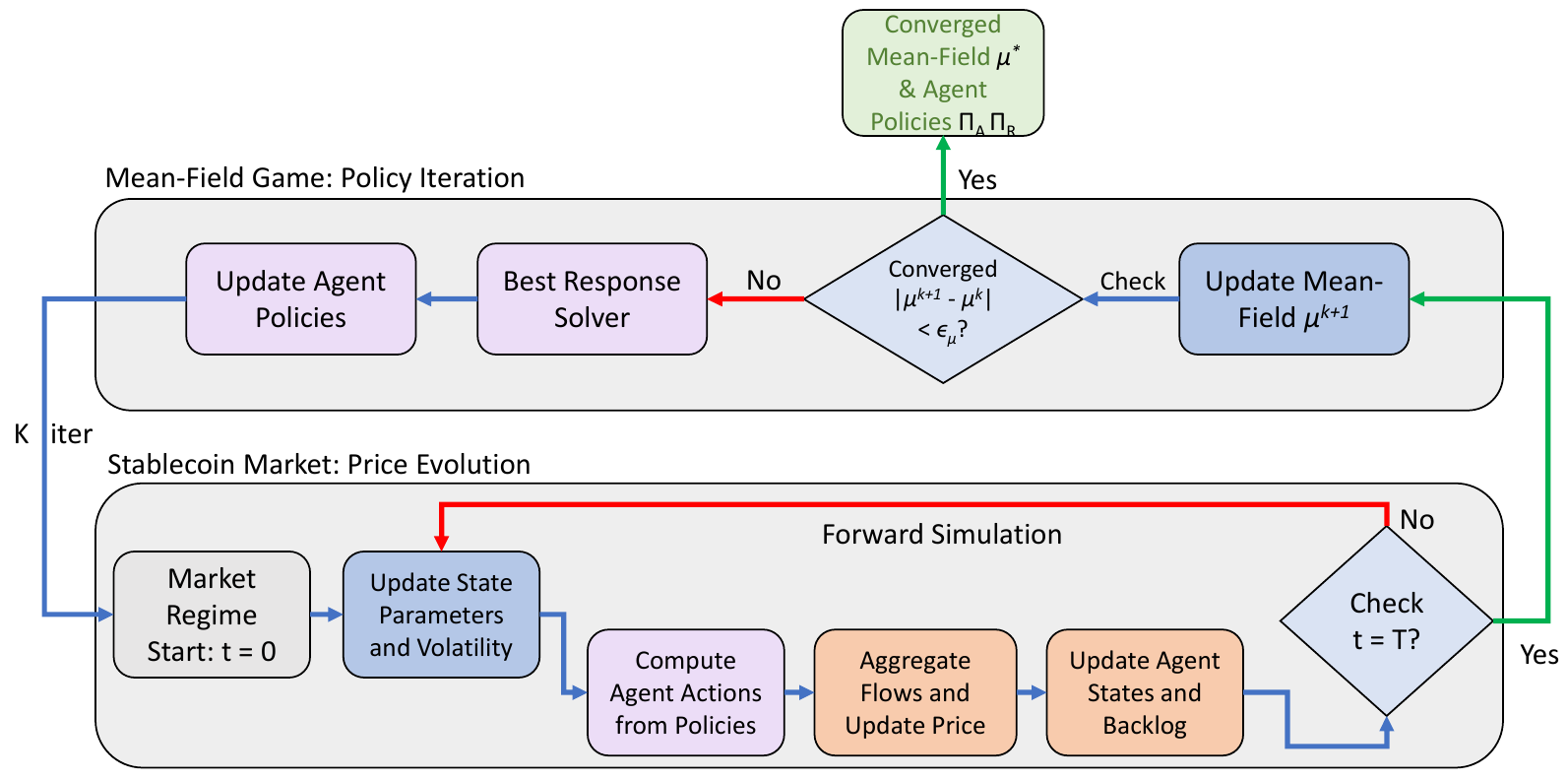}
    \caption{Overview of the Mean-Field Game (MFG) framework used to model agent dynamics for stablecoin markets.}
    \label{fig:mfg_flowchart}
\end{figure}

\section{Related Work}\label{sec:related_work}

\subsection{Stablecoin Peg Arbitrage and Restoration Mechanics}

An increasing body of work studies how stablecoins maintain their pegs and what happens when those pegs come under stress. In a historical and regulatory perspective, Gorton \textit{et al.} \cite{gorton2023taming} interpret fiat-backed stablecoins as a new form of private bank money and argue that, without bank-style regulation and sufficiently safe collateral, they are vulnerable to runs and failure to trade at par. Lyons \textit{et al.} \cite{lyons2023keeps} use detailed data on treasury trades and order books for Tether to show that reforms which broadened access to primary-market arbitrage, such as migration to Ethereum and decentralized issuance, reduced typical peg deviations by roughly half and that premiums and discounts are shaped by safe-haven demand and collateral constraints. Recent empirical work also begins to quantify which design features and market environments are associated with more frequent or severe de-pegs, including the interaction with crypto lending and leverage \cite{makarov2020trading, ante2022liquidity}. However, these empirical contributions primarily estimate reduced-form price dynamics and aggregate arbitrage flows and do not explicitly model the strategic interaction of a large population of retail traders and arbitrageurs across primary and secondary venues.

\subsection{Market Microstructure of Digital Assets}

Our framework explicitly models market frictions, which are now central in the study of digital asset markets. Seminal work by Makarov and Schoar \cite{makarov2020trading} documents large and persistent cross-exchange price deviations and shows that a common component in order flow explains much of the common component in returns across bitcoin exchanges, highlighting both segmentation and non-trivial execution costs. Stablecoin-specific work by Pernice \cite{pernice2021stablecoin} develops a continuous-time model of stablecoin prices and arbitrage bounds under collateral and funding frictions, formalizing how redemption costs and balance sheet constraints shape the no-arbitrage region for fiat-backed coins.
Beyond these early contributions, a growing empirical microstructure literature studies order flow, liquidity and price impact in crypto markets, often with an explicit focus on stablecoin pairs \cite{barucci2023market, cortez2020exchange, cartea2025decentralised}.  
However, these studies generally treat trader behaviour as reduced-form and do not explicitly solve for the optimal, forward-looking policies of a large population of strategic agents, such as retail traders and arbitrageurs, whose collective actions generate the observed order flow and price dynamics.

\subsection{Game-Theoretic Models for Stablecoin Markets}
Game-theoretic frameworks have been increasingly used to understand the fundamental drivers of financial system stability. A prominent example in stablecoin markets is the work by Potter \textit{et al.} \cite{potter2024drives}, which provides a game-theoretical model to identify why stablecoins de-peg. Their model analyzes different price equilibria that emerge based on a coin's underlying architecture and pegging mechanism. This provides a formal basis for comparing the relative price stability of stablecoin designs. 
While prior work studies stablecoin architectures through static design and equilibrium comparisons, our approach instead builds a fully dynamic, agent-based MFG that models how a large population of interacting traders and arbitrageurs respond during a de-peg episode. Our methodological approach draws from MFG theory, introduced by Lasry and Lions \cite{lasry2007mean}, with the specific Linear-Quadratic (LQ) framework being formalized in work such as \cite{bensoussan2016linear, uz2020reinforcement}. 

This framework is particularly suited for analyzing agent dynamics in stablecoin markets and has been successfully applied to models of bank runs \cite{carmona2017mean}, which serve as a strong theoretical analogue to a stablecoin de-peg (a run on reserves). Furthermore, MFGs have been used to model market microstructure, particularly the impact of crowd trading on optimal execution \cite{cardaliaguet2018mean}. More recently, there has been an increasing interest in MFG with the rise of LLM and RL-based agents \cite{lauriere2022learning}. 
To the best of our knowledge, there is currently no dynamic, agent-based MFG model calibrated to observed stablecoin de-peg episodes that explicitly captures the dynamic interaction of multiple strategic agents across primary and secondary markets.

\section{Methodology}\label{sec:methodology}

\subsection{Dynamic Mean-Field Game Framework}
A dynamic MFG models a system with a continuum of rational, non-atomic agents who interact via a ``mean field" \cite{lauriere2022learning}. In our model, as illustrated in Figure~\ref{fig:mfg_flowchart}, this mean field represents the aggregate market state. In our setting, the mean-field state at time $t$ is $\mu_t = (m_t, L_t, \phi_t, \psi_t)$, where $m_t$ is the stablecoin mispricing relative to the one-dollar peg, $L_t$ is the vector of primary backlogs across chains, $\phi_t$ is the vector of aggregate secondary flows across venues, and $\psi_t$ is the vector of aggregate primary flows. Different components of $\mu_t$ enter different parts of the agent cost functions: $m_t$ drives inventory and pricing costs, $L_t$ and $\psi_t$ determine primary access costs, and $\phi_t$ enters the secondary-market congestion term. The net order flow, in turn, determines market-wide variables like price slippage and execution costs. 

\begin{algorithm}[!h]
\small
\caption{MFE Policy Iteration Algorithm}
\label{alg:policy_iteration}
\begin{algorithmic}[1]
  \REQUIRE $(\mathcal{P}, \mathcal{A}, \mathcal{M})$, $(\epsilon_{\text{exploit}}, \epsilon_{\mu}, k_{\max})$
  \STATE Initialize: $k \gets 0$, $\mu^0 \leftarrow \{ m_t, L_t, \phi_t, \psi_t \}_{t=0}^{T}$
  \REPEAT
    \STATE $k \gets k + 1$
    \STATE $\mu^{\text{prev}} \gets \mu^{k-1}$
    \STATE $(\Pi_R, \Pi_A) \gets \textsc{SolveBestResponse}(\mathcal{P}, \mathcal{A}, \mathcal{M}, \mu^{\text{prev}})$
    \STATE $\mu^{k} \gets \textsc{UpdateMeanField}(\mathcal{P}, \mathcal{A}, \mathcal{M}, \Pi_R, \Pi_A)$
    \STATE $\text{Exploit} \gets \textsc{ComputeExploitability}(\mathcal{P}, \mathcal{A}, \mathcal{M}, \mu^{k})$
    \STATE $E \gets \text{Exploit}[\text{max\_exploit}]$
    \STATE $\Delta_{\mu} \gets \lVert \mu^{k}_{\text{price}} - \mu^{\text{prev}}_{\text{price}} \rVert_{\infty}$
  \UNTIL{$(E < \epsilon_{\text{exploit}} \ \land \ \Delta_{\mu} < \epsilon_{\mu}) \ \lor \ k \ge k_{\max}$}
  \STATE \textbf{return} converged mean field $\mu^{*}$ and policies $(\Pi_R, \Pi_A)$
\end{algorithmic}
\end{algorithm}

\begin{figure*}[!h]
    \centering
    \begin{minipage}[t]{0.32\textwidth}
        \centering
        \includegraphics[width=\linewidth]{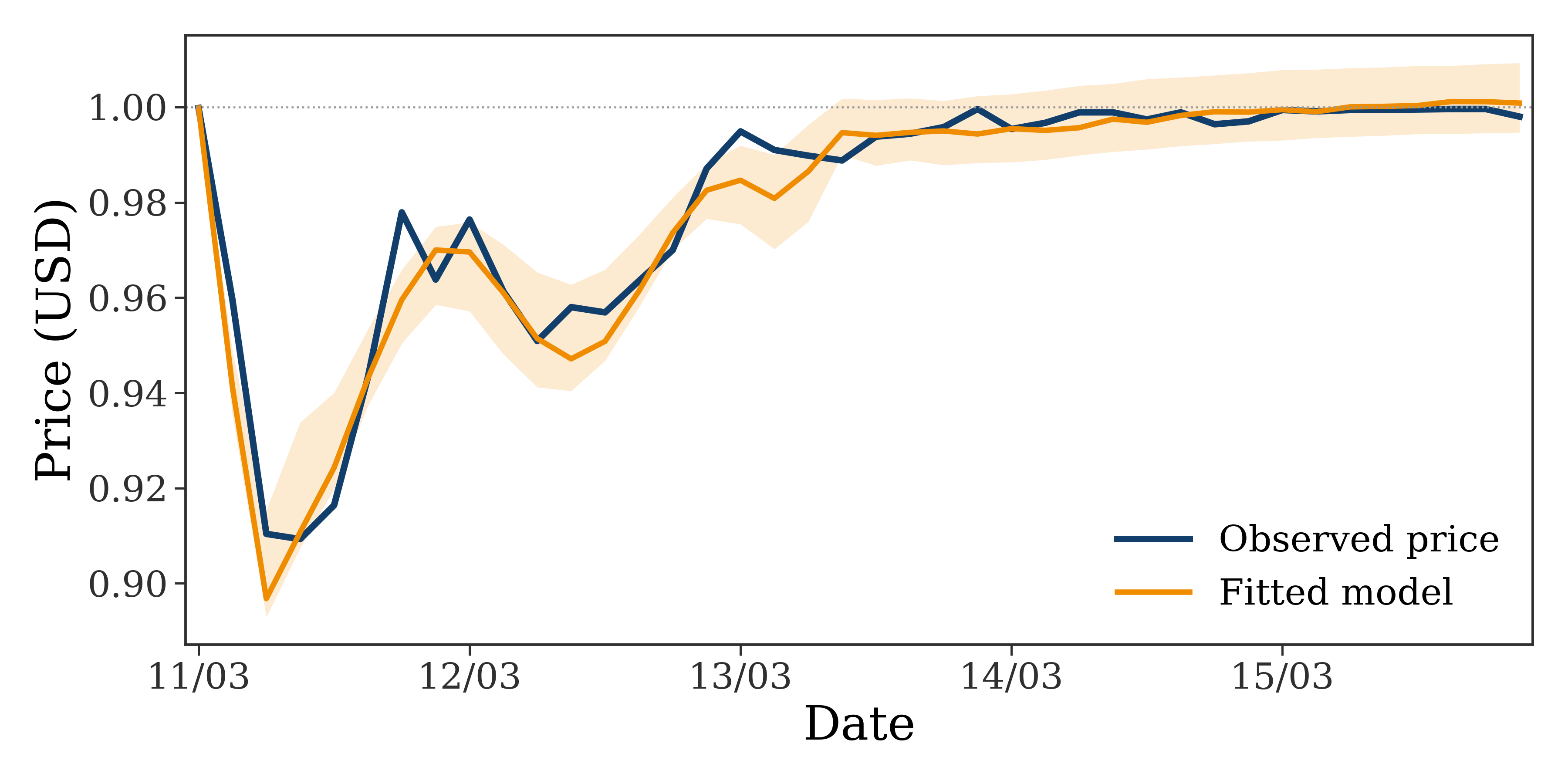}
    \end{minipage}\hfill
    \begin{minipage}[t]{0.32\textwidth}
        \centering
        \includegraphics[width=\linewidth]{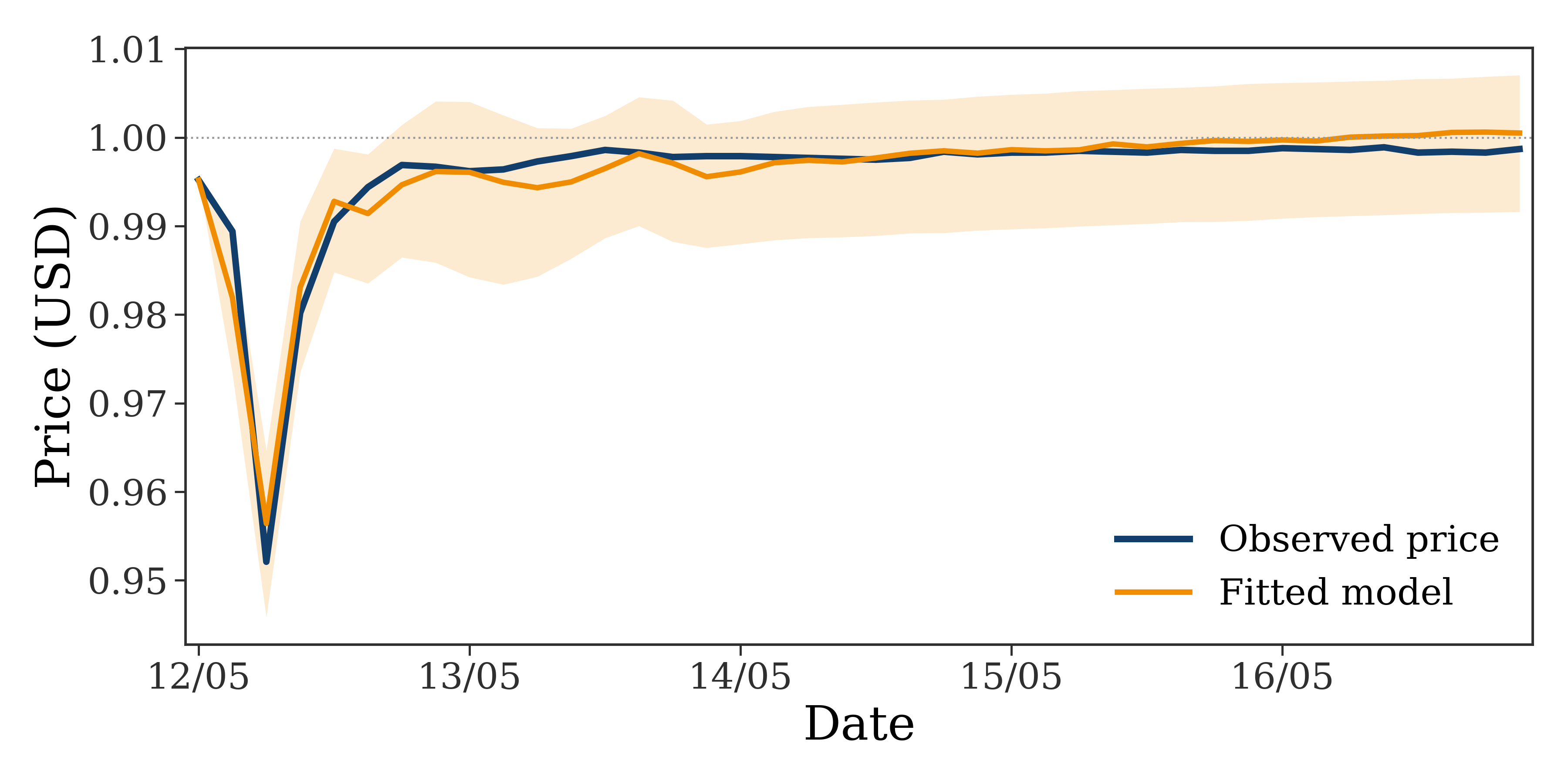}
    \end{minipage}\hfill
    \begin{minipage}[t]{0.32\textwidth}
        \centering
        \includegraphics[width=\linewidth]{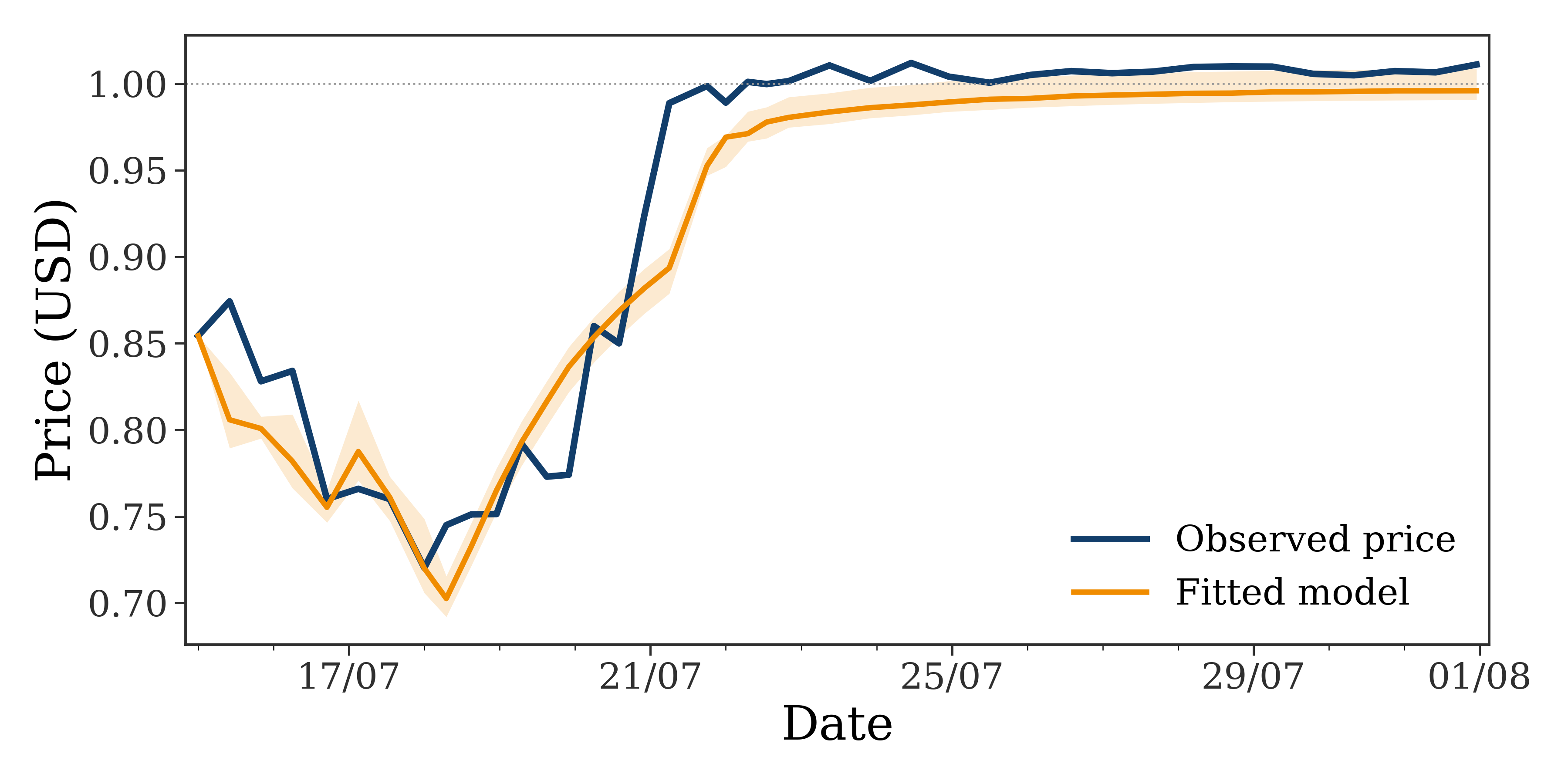}
    \end{minipage}
    \caption{Comparison of the simulated stablecoin price path against the historical observed price data during each de-peg event.}
    \label{fig:stablecoin_depeg_fit}
\end{figure*}

\subsection{Economic Environment and Agent Types}
Our model consists of two main components: a multi-population agent system and a multi-venue market microstructure. The market is populated by a continuum of agents, divided into two classes (1) Retail Traders: A fraction $\pi_R$ of the population, characterized by higher execution frictions such as slippage costs $\kappa_R$ and inventory aversion $\eta_R$. They are assumed to trade exclusively in secondary markets. Their state is their inventory $q_{R,t}$ and their control variable is their vector of secondary market flows $a_{R,t} \in \mathbb{R}^S$. (2) Arbitrageurs: A fraction $\pi_A$ of the population, representing sophisticated actors with lower secondary market frictions $\kappa_A$ and inventory aversion $\eta_A$. They have access to both primary and secondary markets. Their state is their inventory $q_{A,t}$ and their control variables are secondary market flows $a_{A,t} \in \mathbb{R}^S$ and primary redemption flows $r_t \in \mathbb{R}^C$.

Both agent types solve a discounted LQ optimization problem to minimize a cost functional that combines mispricing exposure, inventory risk, execution costs, and congestion costs. The first term in each objective corresponds to the negative of trading profit from exploiting deviations from the peg, while the remaining terms are cost penalties.
Quadratic execution costs of this type are standard in optimal execution and mean-field trading models, where linear temporary price impact in trade size implies a strictly convex cost term \cite{almgren2003optimal, cartea2025decentralised}. 
An agent of type $i \in \{R, A\}$ chooses control policies (flows) to minimize an expected sum of discounted future costs with discount factor $\gamma \in (0,1)$. 
In particular, retail traders solve:
\begin{equation}
\label{eq:retail_obj}
\begin{aligned}
\min_{\{a_{R,t}\}} \quad
& \mathbb{E} \Bigg[
    \sum_{t=0}^{\infty} \gamma^t \Big(
        \underbrace{m_t \sum_{s=1}^S a_{R,s,t}}_{\text{Trading PnL}}
        + \underbrace{\frac{1}{2}\eta_R q_{R,t}^2}_{\text{Inventory Cost}} \\
& \qquad\qquad
        + \underbrace{\frac{1}{2} \sum_{s=1}^S \kappa_{R,s} a_{R,s,t}^2}_{\text{Execution Cost}}
        + \underbrace{C(a_{R,t}, \bar{a}_{R,t})}_{\text{Congestion Cost}}
    \Big)
\Bigg] \\
\text{s.t.} \quad
& q_{R,t+1} = q_{R,t} + \sum_{s=1}^S a_{R,s,t} \, .
\end{aligned}
\end{equation}
Here $C(a_{R,t}, \bar a_{R,t}) = \xi_R\, a_{R,t}^\top \phi_t$ denotes a mean-field congestion cost in secondary markets, which we implement using the aggregate secondary flow $\phi_t$, the secondary-flow component of the mean-field state $\mu_t$.

Arbitrageurs solve an analogous control problem that is augmented to account for both secondary and primary flows. Let $\tau_c(L_t)$ denote the effective linear cost of routing one unit through primary chain $c$, which depends on the backlog state $L_t$ and the settlement parameters of the protocol. Then, the arbitrageur control problem is
\begin{equation}
\label{eq:arb_obj}
\begin{aligned}
\min_{\{a_{A,t}, r_t\}} \quad
& \mathbb{E} \Bigg[
    \sum_{t=0}^{\infty} \gamma^t \Big(
        \underbrace{m_t \sum_{s=1}^S a_{A,s,t}}_{\text{Secondary Trading PnL}}
        + \underbrace{\sum_{c=1}^C \tau_c(L_t)\, r_{c,t}}_{\text{Primary Routing Cost}} +\\
&       \underbrace{\cdots}_{\text{Secondary Costs}} 
        + \underbrace{\frac{1}{2} \sum_{c=1}^C \kappa_{P,c} r_{c,t}^2}_{\text{Primary Execution Cost}}
        + \underbrace{C_A(a_{A,t}, \bar{a}_{A,t})}_{\text{Congestion Cost}}
    \Big)
\Bigg] \\
\text{s.t.} \quad
& q_{A,t+1} = q_{A,t} + \sum_{s=1}^S a_{A,s,t} + \sum_{c=1}^C r_{c,t} \, .
\end{aligned}
\end{equation}
The term $C_A(a_{A,t}, \bar a_{A,t}) = \xi_A\, a_{A,t}^\top \phi_t$ denotes the congestion cost for arbitrageurs, parameterized by a coefficient $\xi_A$. In the special case $\xi_R = \xi_A = 0$, the congestion terms vanish and both \eqref{eq:retail_obj} and \eqref{eq:arb_obj} reduce to standard discounted LQ control problems with linear state dynamics.

\subsection{Stablecoin Market Dynamics}
The market environment and mean fields (mispricing, backlog, aggregate primary and secondary flows) evolve based on the aggregate actions of all agents. The secondary and primary markets consist of a set of $S$ distinct venues (e.g., CEX, DEX) with heterogeneous price impact parameters $\lambda_s$ and a set of $C$ mint/redeem channels (different blockchains) that allow arbitrageurs to mint or redeem stablecoins at the \$1 peg, subject to costs $\kappa_P$. The secondary market price deviation ($m_t$) evolves based on the net order flow from all agents and an exogenous shock $\epsilon_t$ as follows:
\begin{equation}
\label{eq:price_dyn}
\begin{aligned}
m_{t+1}
= {} & m_t
    + \sum_{s=1}^S \lambda_s \bigl( \pi_R a_{R,s,t} + \pi_A a_{A,s,t} \bigr) \\
& {} + \sum_{c=1}^C \gamma_c \bigl( \pi_A r_{c,t} \bigr)
    + \epsilon_t .
\end{aligned}
\end{equation}
where, $\gamma_c$ is the parameter coupling primary flows to the secondary market price. The backlog for each channel $c$ is represented by $L_{c,t}$. It evolves based on a processing rate $\delta_c$ (decay) and new arbitrageur submissions. It represents a mean-field state tracking redemption submissions
\begin{equation}
    L_{c,t+1} = (1-\delta_c)L_{c,t} + \pi_Ar_{c,t}
\end{equation}

\subsection{Stochastic Volatility and State-Dependent Parameters}
To capture realistic, time-varying market stress, the market model incorporates state-dependent dynamics with a stochastic volatility process. The volatility is modeled as a GARCH process \cite{bollerslev1986generalized} for the conditional variance $\sigma^2_t$ of the exogenous price shock $\epsilon_t$. The exogenous shock from Equation~\ref{eq:price_dyn} is defined as $\epsilon_t = \sigma_t Z_t$, where, $Z_t \sim \mathcal{N}(0,1)$ is a standard normal random variable, and the variance evolves according to:
\begin{equation}
    \sigma_t^2 = \omega + \alpha \epsilon_{t-1}^2 + \beta \sigma_{t-1}^2
\end{equation}
where $\omega, \alpha, \beta$ are the standard GARCH parameters calibrated to stablecoin market data.
This time-varying volatility $\sigma_t$ directly modulates key parameters of the market environment and agent costs. Agents' execution frictions ($\kappa$) and inventory aversion ($\eta$) are scaled by the current volatility, reflecting increased risk aversion and costs in turbulent markets
\begin{equation}
\begin{aligned}
    \kappa_{i,s,t} &= \kappa_{i,s,0} (1 + c_i \sigma_t) \\
    \eta_{i,t} &= \eta_{i,0} (1 + d_i \sigma_t)
\end{aligned}
\end{equation}
where $c_i$ and $d_i$ are the risk-scaling sensitivities for agent type $i \in \{R, A\}$. Secondary market price impact (liquidity) also becomes volatility-dependent, with each venue $s$ having its own sensitivity $a_s$
\begin{equation}
    \lambda_{s,t} = \lambda_{s,0} (1 + a_s \sigma_t)
\end{equation}
Agents employ a softmax-based routing logic to dynamically allocate trades across secondary venues, minimizing their effective, time-varying costs which depend on the now state-dependent impact $\lambda_{s,t}$ and frictions $\kappa_{i,s,t}$.

\begin{figure*}[!h]
    \centering
    \begin{minipage}[t]{0.32\textwidth}
        \centering
        \includegraphics[width=\linewidth]{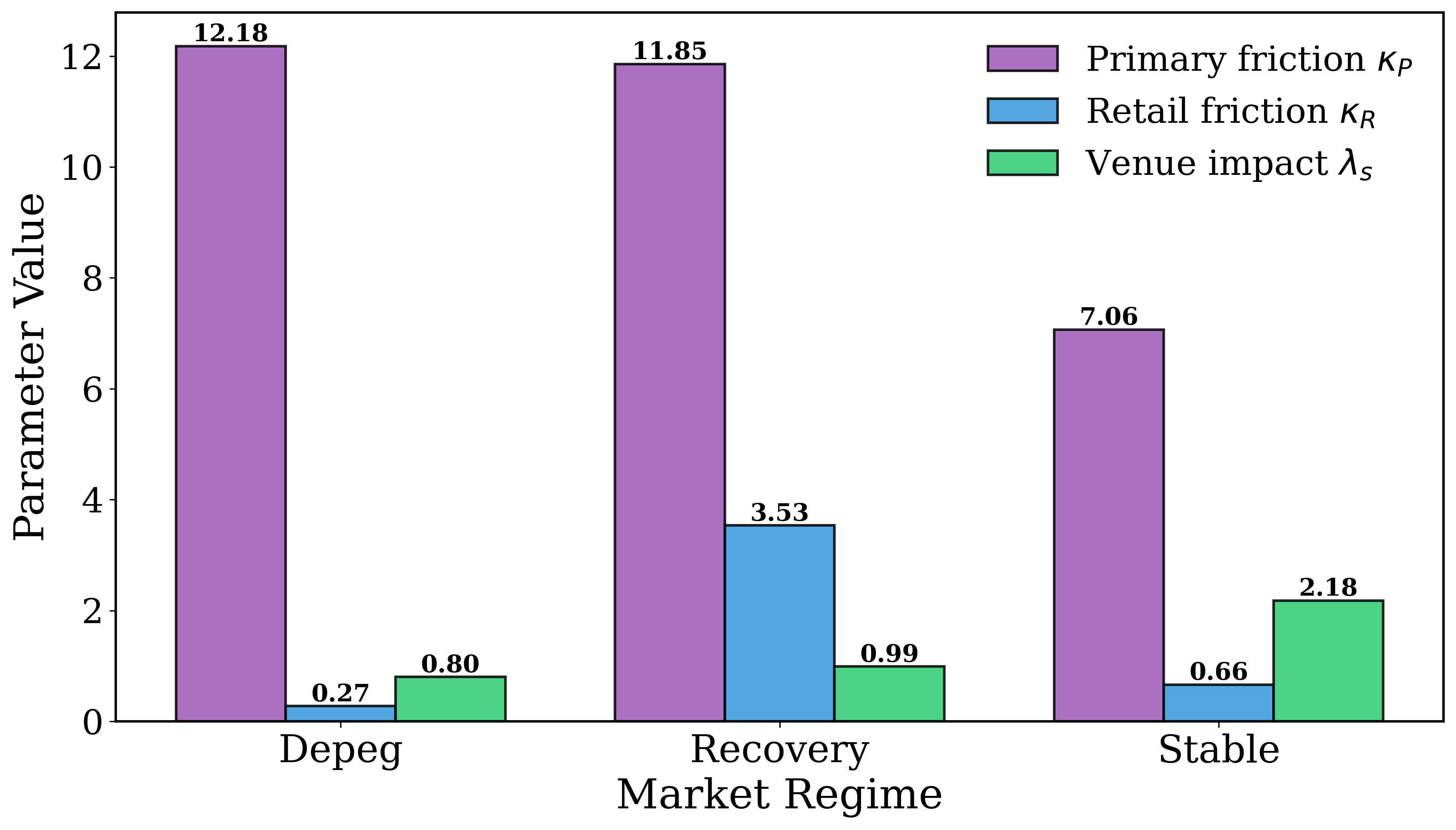}
    \end{minipage}\hfill
    \begin{minipage}[t]{0.32\textwidth}
        \centering
        \includegraphics[width=\linewidth]{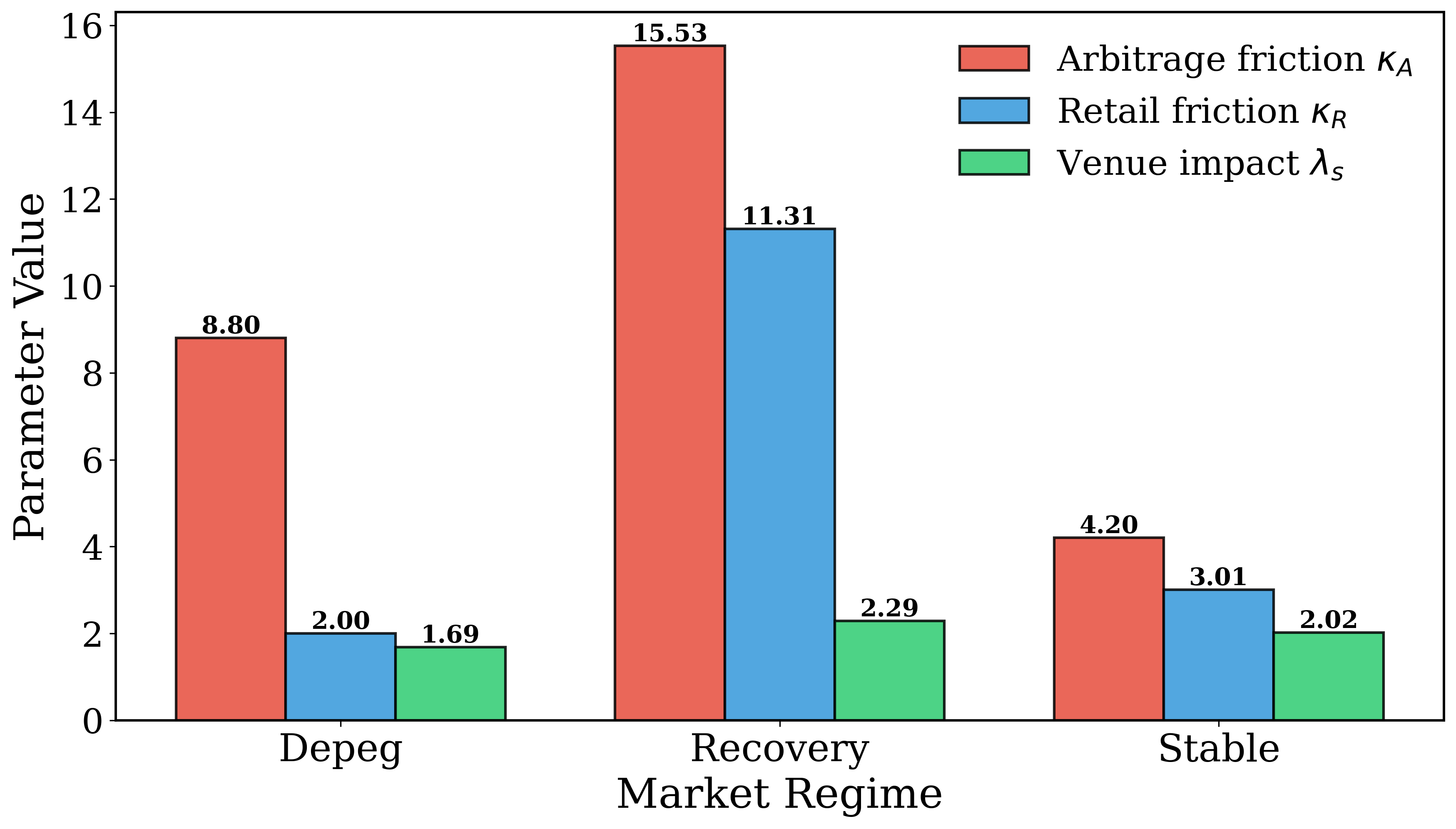}
    \end{minipage}\hfill
    \begin{minipage}[t]{0.32\textwidth}
        \centering
        \includegraphics[width=\linewidth]{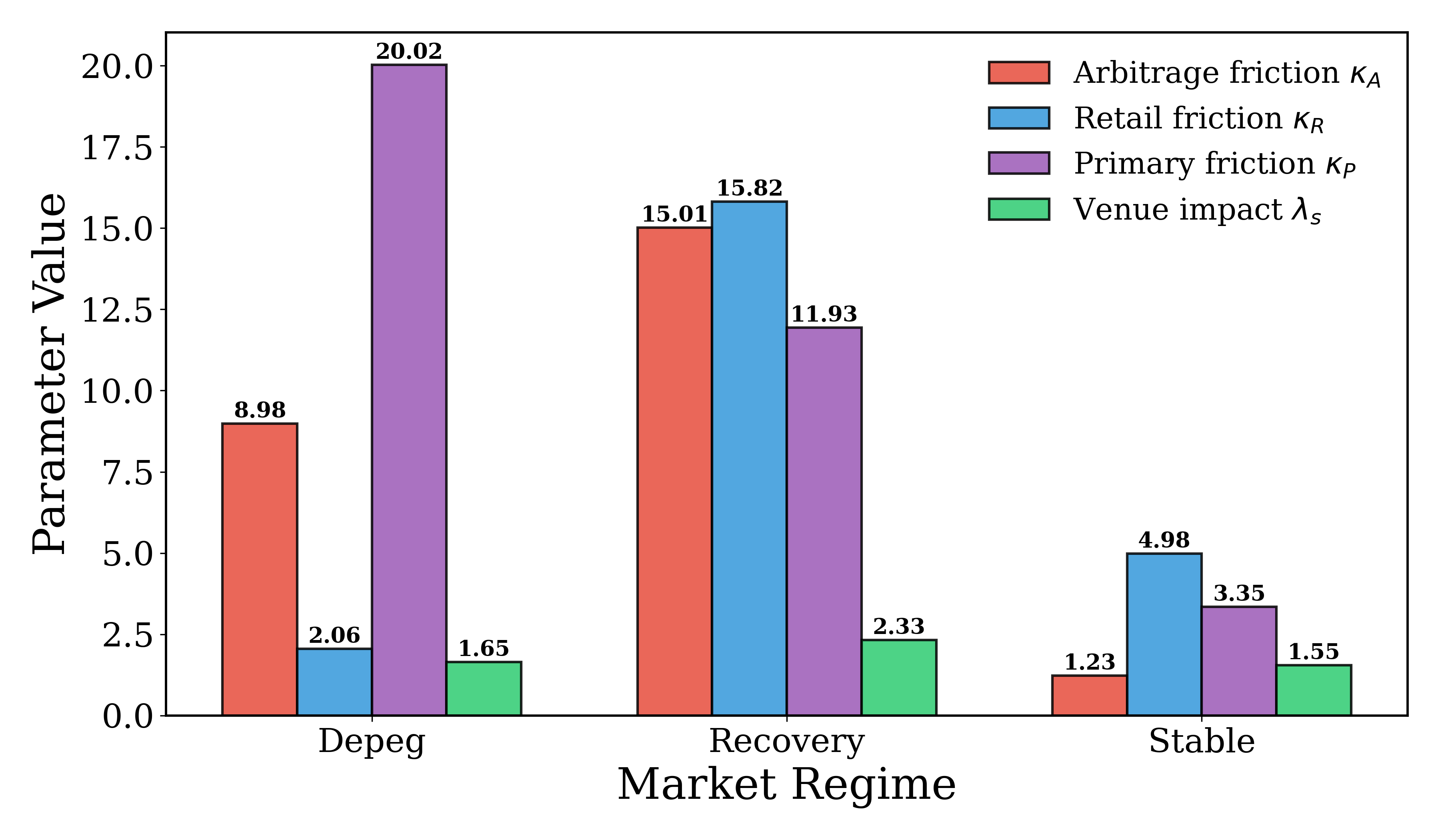}
    \end{minipage}
    \caption{Calibrated model parameters across three distinct market regimes identified during each de-peg event.}
    \label{fig:stablecoin_depeg_param}
\end{figure*}

\subsection{Mean-Field Equilibrium Computation}
We solve for the Mean-Field Equilibrium (MFE) computationally. The MFE is a fixed point where the agents' policies are optimal given the mean field. Also, the mean field is consistent with the aggregation of those policies.
Our approach employs the Policy Iteration algorithm, as detailed in Algorithm~\ref{alg:policy_iteration}, in line with numerical methods for MFGs and learning-based approaches \cite{lauriere2022learning, cacace2021policy}. The algorithm iteratively computes agents' best-response policies and updates the mean-field dynamics until a fixed point is reached.

The algorithm terminates when two conditions are met: 1) the mean-field trajectory converges, and 2) the system reaches an $\epsilon$-Nash equilibrium. We validate the $\epsilon$-Nash condition by calculating the exploitability ($E$), which measures the maximum profit any agent could gain by deviating from the MFE policy. By ensuring $E < \epsilon_{exploit}$, we confirm that no agent has a significant incentive to deviate from the equilibrium policy. 

\begin{table}
\centering
\caption{Baseline Model Calibrated Parameters}
\label{tab:baseline_params}
\begin{tabular}{@{}lll@{}}
\toprule
Category & Parameter & Value \\
\midrule
Simulation Control & Time horizon ($T$) & $40$ \\
& Discount factor ($\gamma$) & $0.97$ \\
& Initial mispricing ($m_0$) & $-0.01$ \\
& Time step ($\Delta t$) & $1.0$ \\
\addlinespace
Market Structure & Secondary venues ($S$) & $3$ \\
& Primary channels ($C$) & $2$ \\
& Secondary price impact ($\lambda_s$) & $[1.6, 1.8, 2.5]$ \\
& Primary price impact ($\gamma_c$) & $[2.0, 1.4]$ \\
& Venue weights ($w$) & $[0.5, 0.35, 0.15]$ \\
\addlinespace
Agent Population & Retail share ($\pi_R$) & $0.85$ \\
& Arbitrageur share ($\pi_A$) & $0.15$ \\
\addlinespace
Retail Costs & Secondary execution ($\kappa_R$) & $[2.0, 3.0, 4.0]$ \\
& Inventory aversion ($\eta_R$) & $0.15$ \\
& Congestion cost ($\xi_R$) & $0.5$ \\
\addlinespace
Arbitrageur Costs & Secondary execution ($\kappa_A$) & $[1.2, 1.5, 2.0]$ \\
& Primary execution ($\kappa_P$) & $[0.8, 0.6]$ \\
& Inventory aversion ($\eta_A$) & $0.20$ \\
& Congestion cost ($\xi_A$) & $0.3$ \\
\bottomrule
\end{tabular}
\end{table}

\subsection{Model Calibration and Parameter Estimation}

We calibrate our model by testing it against real-world de-peg events. We do this in two steps: first, we split the event data into phases, and second, we estimate the model's parameters.
We analyze historical price data by splitting each de-peg event into three distinct, consecutive phases. This helps us see how market behavior changes over time. We define the phases as:
\begin{itemize}
    \item De-peg Phase: This period starts when the price first drops sharply below \$1. It's marked by high volatility, falling prices, and market stress.
    \item Recovery Phase: This phase starts from the lowest price (de-peg). It covers the time when the price is clearly climbing back toward \$1.
    \item Stable Phase: This is the final phase where the price has returned to \$1 and stays within a normal, tight range.
\end{itemize}

Market data does not directly reveal many key parameters of the model. We therefore estimate them by fitting the model's price predictions to observed price dynamics. Following Richiardi \textit{et al.}~\cite{richiardi2006common}, we treat calibration as a systematic exploration of the admissible parameter space rather than an informal tuning exercise. This full exploration principle is particularly important in agent-based models, where non-linear interactions can generate sharp regime changes and multiple locally optimal parameter regions.

We measure the accuracy of the price fit using a score, $\mathcal{L}(\theta)$, such as mean squared error (MSE). We find the best parameters, $\theta^*$, by finding the set that gives the lowest MSE:
\begin{equation}
\theta^* = \arg \min_{\theta} \mathcal{L}(\theta) = \arg \min_{\theta} \left[ \frac{1}{T} \sum_{t=1}^{T} ( m_t(\theta) - m_t^{\text{obs}} )^2 \right]
\end{equation}
where $m_t(\theta)$ is model-implied mispricing and $m_t^{\text{obs}}$ is the empirical mispricing. We solve this non-convex problem using Differential Evolution \cite{storn1997differential}, a population-based global optimizer that improves robustness to local minima.

\section{Simulation Parameters}\label{sec:simulation_parameters}

\subsection{Baseline Calibration}
To analyze the model, we first establish a baseline calibration designed to realistically reflect the normal market conditions for fiat-collateralized stablecoins as shown in Table~\ref{tab:baseline_params}. 
For the dynamic environment, we work with a discrete time horizon of $T = 40$ periods and a per–period discount factor of $\gamma = 0.97$, which corresponds to the horizon and discounting used in the calibrated stable regime.

\subsection{Dataset Extraction}
For the empirical analysis, we use high–frequency spot price data for the stablecoins USDC and USDT from Binance \cite{binance_us_2025}. We extract Kline (candlestick) data from Binance's public spot market API, which reports open, high, low, close, and traded volume for fixed time intervals. 
For the March 2023 USDC de-peg, we use one–minute Klines for USDC/USD on Binance, covering the period from March 11, 2023, to March 15, 2023 (UTC). This window covers the announcement of Silicon Valley Bank’s receivership, the sharp weekend discount in USDC, and the subsequent recovery once primary redemption channels and off–chain reserve access were restored. 

For the May 2022 USDT de-peg linked to the TerraUSD collapse, we extract one–minute Klines for USDT/USD trading pair on Binance from May 12, 2022, to May 16, 2022 (UTC). 
Finally, for the localized July 2023 USDT dislocation on Binance, we utilize Kline data for the relevant USDT/USD trading pair from July 15, 2023, to July 31, 2023 (UTC).

\section{Experimental Results}\label{sec:experimental_results}

\subsection{Calibration to De-Peg Events}
To validate the model against significant real-world stress events, we fit the dynamic MFG framework to historical data from three de-peg events. Code is publicly available in \cite{stablecoinpegmfg2026}. Figure~\ref{fig:stablecoin_depeg_fit} provides a direct comparison of the model's performance, as it plots the observed historical price against the price path generated by the fitted model. The model captures the key dynamics of the event, including the sharp initial de-peg, high volatility, and the subsequent recovery path back to the peg. We achieve this fit by calibrating the model to reflect the different market regimes of the event. 

To give intuition for the calibration, Figure~\ref{fig:stablecoin_depeg_param} reports the event-specific values of the key execution and liquidity parameters: primary redemption friction $\kappa_P$, secondary execution frictions (retail $\kappa_R$ and arbitrageur $\kappa_A$), and venue impact $\lambda_s$. These coefficients enter the quadratic execution and price-impact penalties in the agents' objectives, so larger values correspond to more expensive, slower, or capacity-constrained trading per unit notional. Across all three events, the calibrated $\kappa_P$ is highest in stressed regimes and declines toward the stable regime, consistent with temporary impairment or congestion of mint/redeem rails during acute stress. Secondary frictions $\kappa_R$ (and $\kappa_A$) tend to peak in recovery, when order books are thinnest and marginal execution costs are elevated as markets rebalance. The venue-impact terms $\lambda_s$ move in the same direction, rising in stressed regimes when a given net flow produces a larger price response. Importantly, these are effective parameters estimated by minimizing the price-path loss over each regime window, so they should be interpreted as capturing the composite of fees, slippage, balance-sheet constraints, and operational delays in that episode. These figures demonstrate the model's ability to replicate a complex, real-world de-peg by dynamically adjusting parameters to reflect underlying market disruptions, such as primary rail failures and the resulting congestion.

\begin{figure}[!t]
    \centering
    \includegraphics[width=0.8\columnwidth]{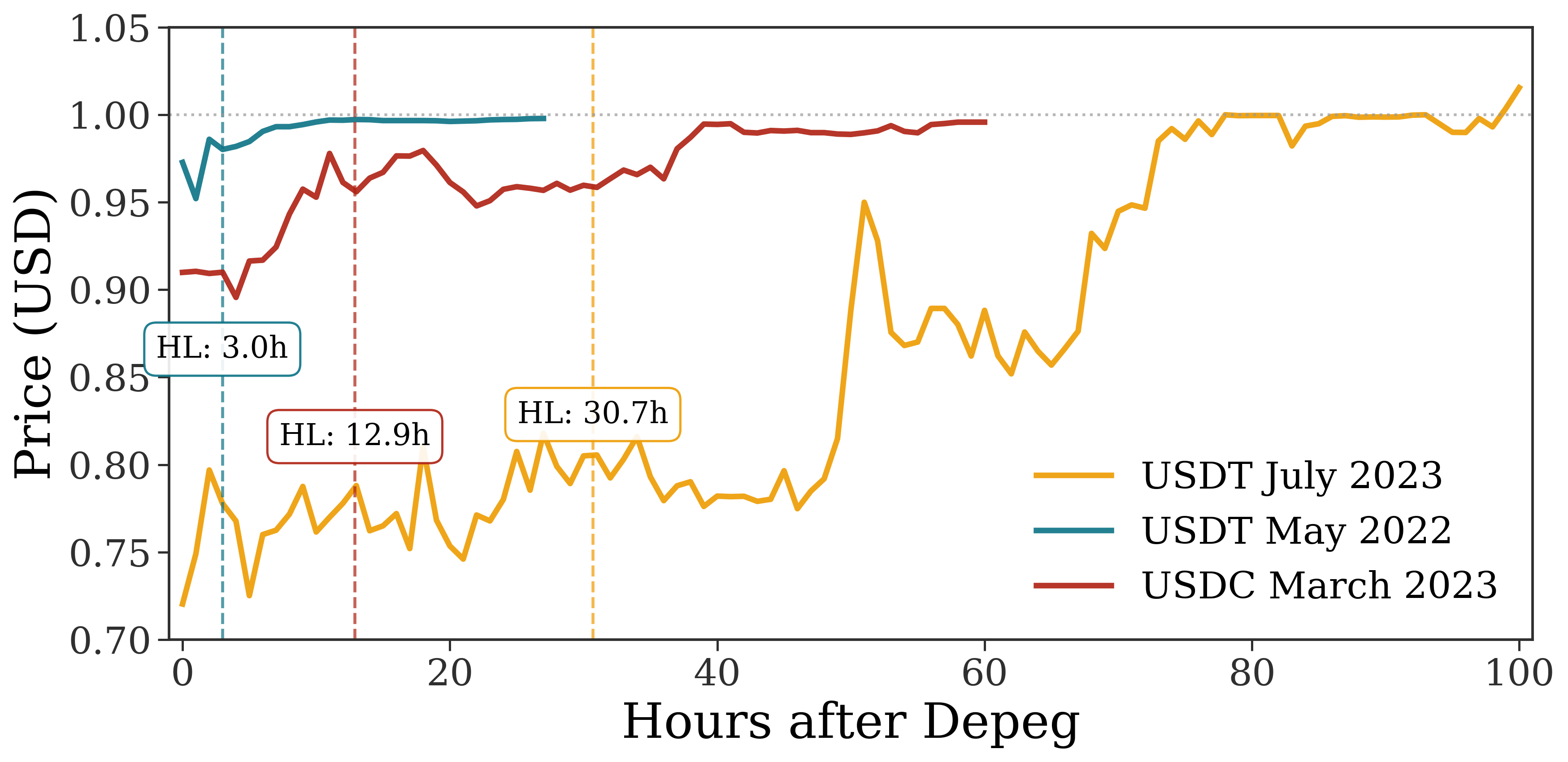}
    \caption{Historical price trajectories and estimated AR(1) half-lives for the three de-peg events.}
    \label{fig:half_life_compare}
\end{figure}

\begin{figure*}[!h]
    \centering
    \begin{minipage}[t]{0.32\textwidth}
        \centering
        \includegraphics[width=\linewidth]{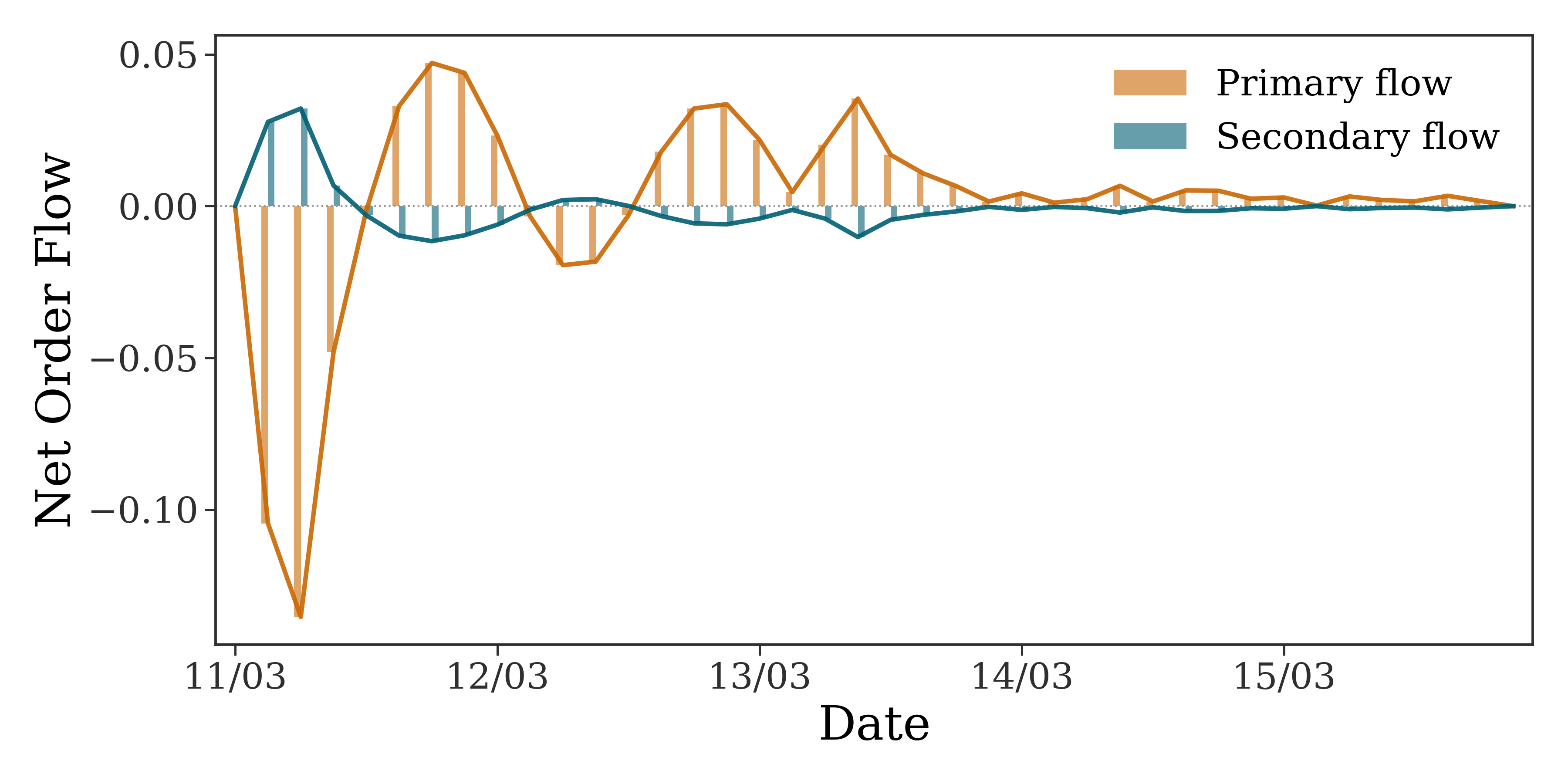}
    \end{minipage}\hfill
    \begin{minipage}[t]{0.32\textwidth}
        \centering
        \includegraphics[width=\linewidth]{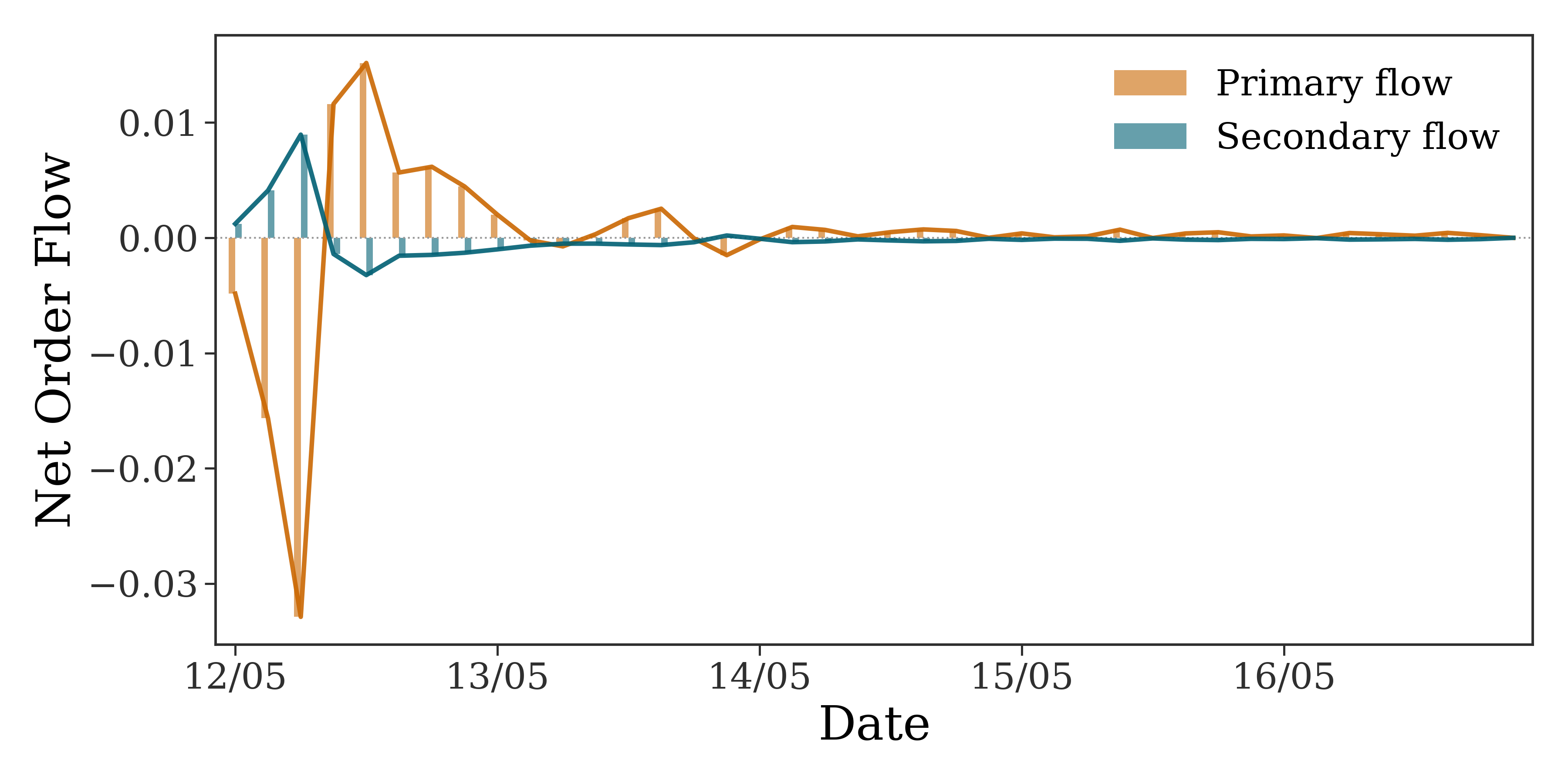}
    \end{minipage}\hfill
    \begin{minipage}[t]{0.32\textwidth}
        \centering
        \includegraphics[width=\linewidth]{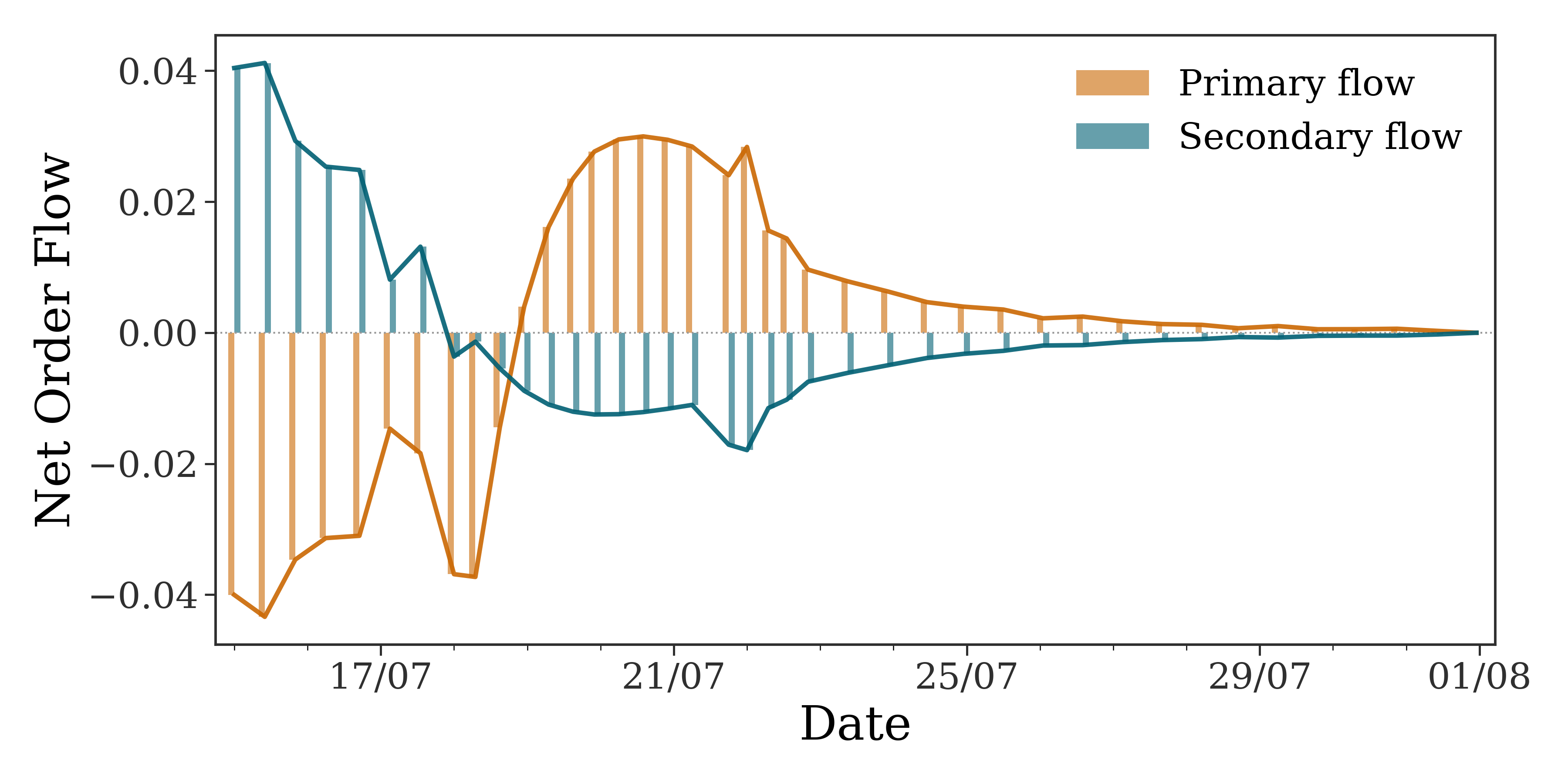}
    \end{minipage}
    \caption{Decomposition of net peg-reverting flows into primary-market mint/redeem flow and secondary-market buying flow on exchanges during each de-peg event. Left: USDC March 2023. Middle: USDT May 2022. Right: USDT July 2023.}
    \label{fig:peg_flow_contrib}
\end{figure*}

\subsection{Estimating the Half-life of De-peg Events}

To quantify the peg recovery speed, we estimate the empirical half-life for each de-peg event and compare this metric to the half-life generated by our fitted MFG model. We define the half-life as the time required for the de-peg to decay by 50\%. To estimate this from historical data, we follow the autoregressive AR(1) methodology \cite{lyons2023keeps}. We fit the mispricing $m_t$ to the process $m_t = \rho m_{t-1} + \epsilon_t$ and calculate the half-life $HL = \ln(2) / (-\ln \rho)$.
Figure~\ref{fig:half_life_compare} plots the historical price trajectories for the three events, with the empirically estimated half-life for each marked by a vertical dashed line. 
For the May 2022 event, both data and model agree on a rapid, sub-3-hour recovery. For the March 2023 event, the model estimates a half-day recovery. For the July 2023 event, the model predicts a 45.9-hour half-life, slightly longer than the 30.7-hour empirical value, but correctly identifies the slow recovery. This result validates that our model, calibrated on agent-level frictions, can translate the micro-level parameters into correct macro-level recovery dynamics.

\begin{table}[t]
    \centering
    \caption{Average performance of the dynamic MFG and reduced-form baselines across the three calibrated de-peg events.}
    \label{tab:baseline_summary}
    \setlength{\tabcolsep}{4pt}
    \begin{tabular}{lccc}
        \toprule
        Model & Avg.\ RMSE & \begin{tabular}{@{}c@{}}Avg.\ half-life error\end{tabular} & \begin{tabular}{@{}c@{}}Captures agent \\ dynamics?\end{tabular} \\
        \midrule
        AR(1)        & 0.0158      & 0.36                  & No \\
        ARMA-GARCH   & 0.0162      & 0.24                  & No \\
        ARX          & 0.0217      & 1.15                  & No \\
        Dynamic MFG  & 0.0128      & 0.41                  & Yes \\
        \bottomrule
    \end{tabular}
\end{table}

\begin{table}[t]
    \centering
    \caption{Out-of-sample performance of the dynamic MFG on the March 2023 USDC de-peg event under different train/test splits.}
    \label{tab:usdc_oos}
    \setlength{\tabcolsep}{4pt}
    \footnotesize
    \begin{tabular}{lcccc}
        \toprule
        Split ratio & Train RMSE & Test RMSE & Train MAE & Test MAE \\
        \midrule
        70/30 & 0.002809 & 0.007234 & 0.001754 & 0.006180 \\
        80/20 & 0.002287 & 0.007048 & 0.001571 & 0.005682 \\
        90/10 & 0.002029 & 0.007272 & 0.001329 & 0.005893 \\
        \bottomrule
    \end{tabular}
\end{table}

\begin{figure}[!h]
    \centering
    \begin{minipage}[t]{0.24\textwidth}
        \centering
        \includegraphics[width=\linewidth]{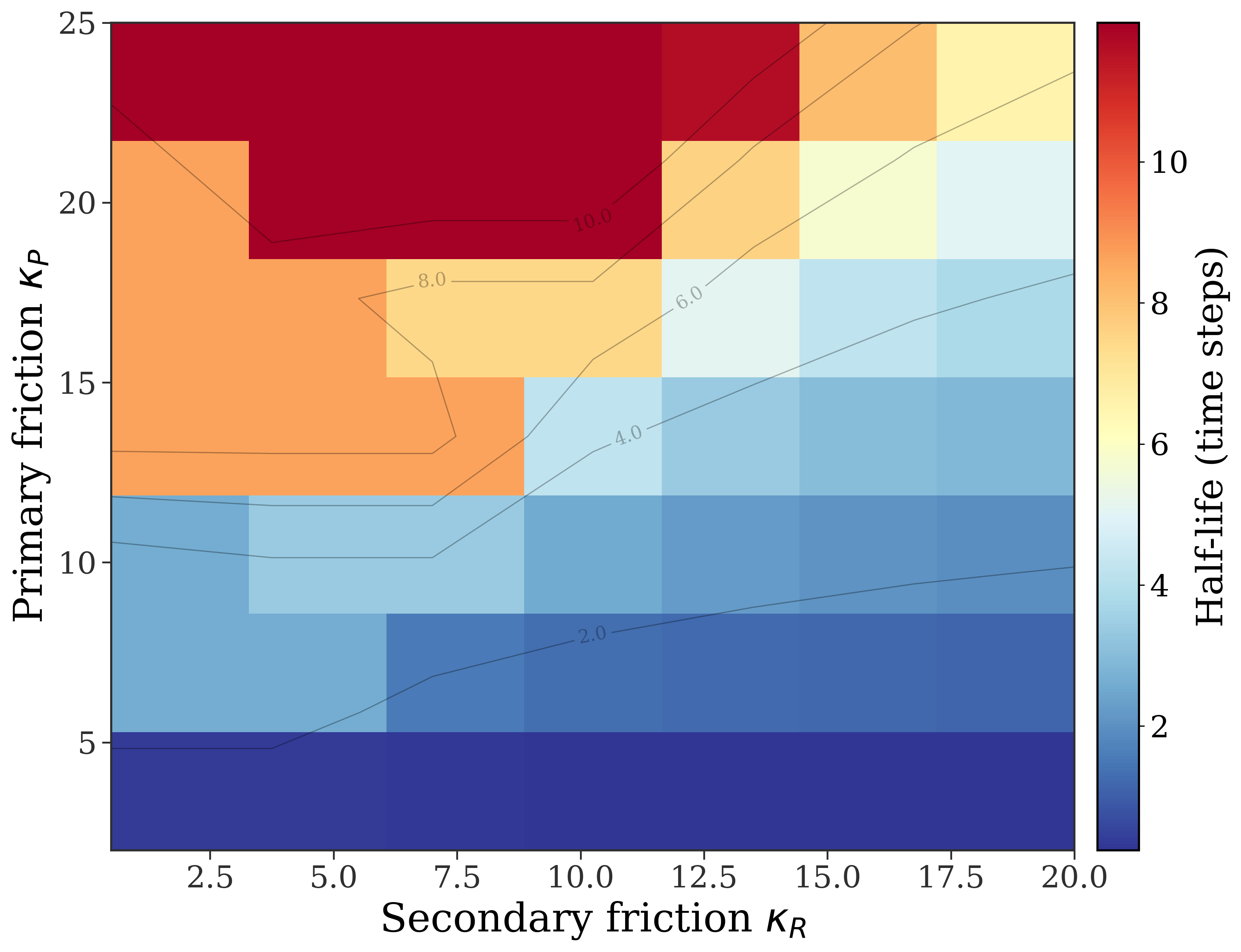}
        \label{fig:heatmap_kA_kR}
    \end{minipage}\hfill
    \begin{minipage}[t]{0.24\textwidth}
        \centering
        \includegraphics[width=\linewidth]{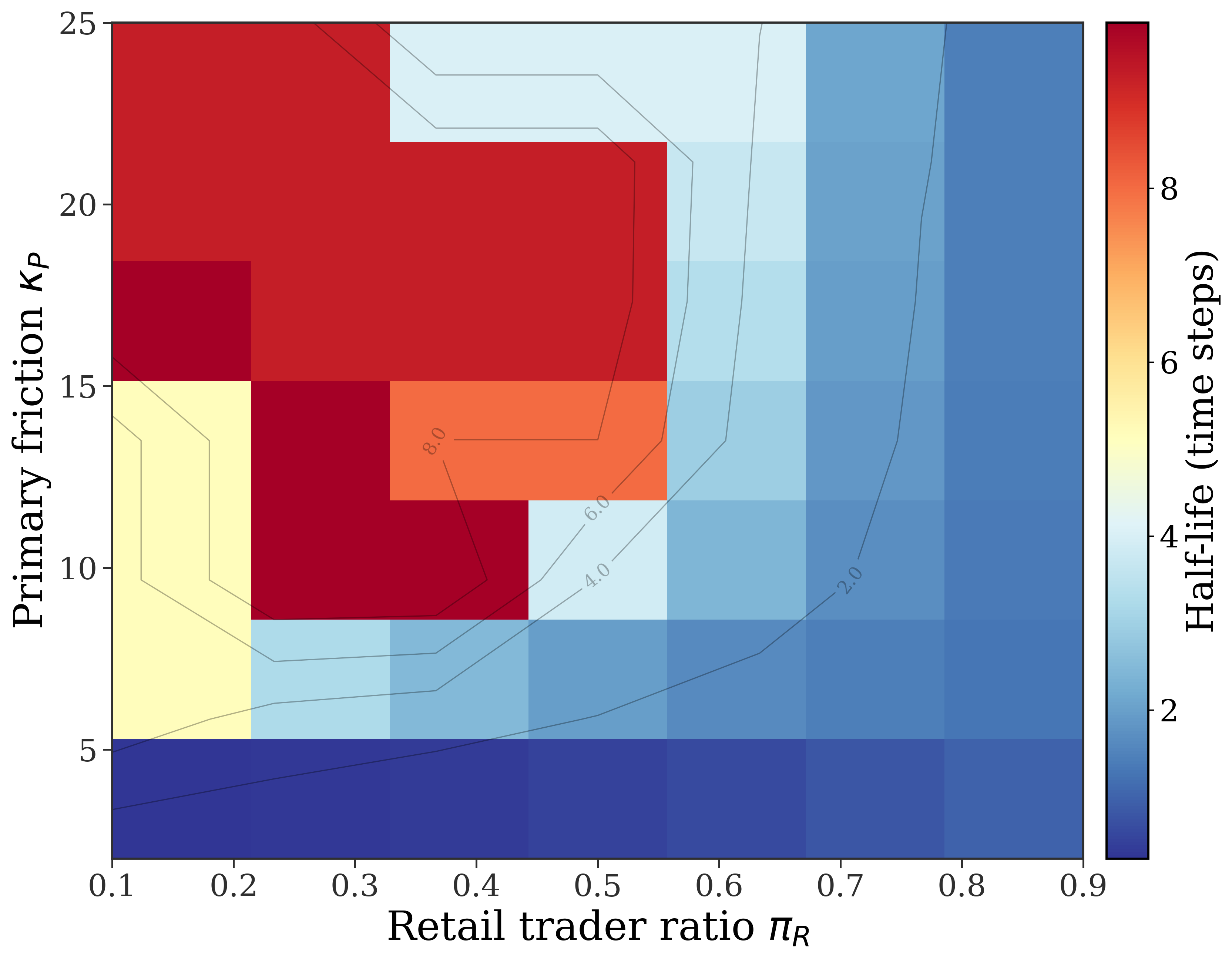}
        \label{fig:heatmap_kA_piR}
    \end{minipage}
    \caption{De-peg half-life sensitivity to market frictions $\kappa_P, \kappa_R$ and composition $\pi_R$.}
    \label{fig:heatmaps_all}
\end{figure}

\subsection{Baselines and Model Comparison}

\begin{figure}[!h]
    \centering
    \includegraphics[width=0.7\columnwidth]{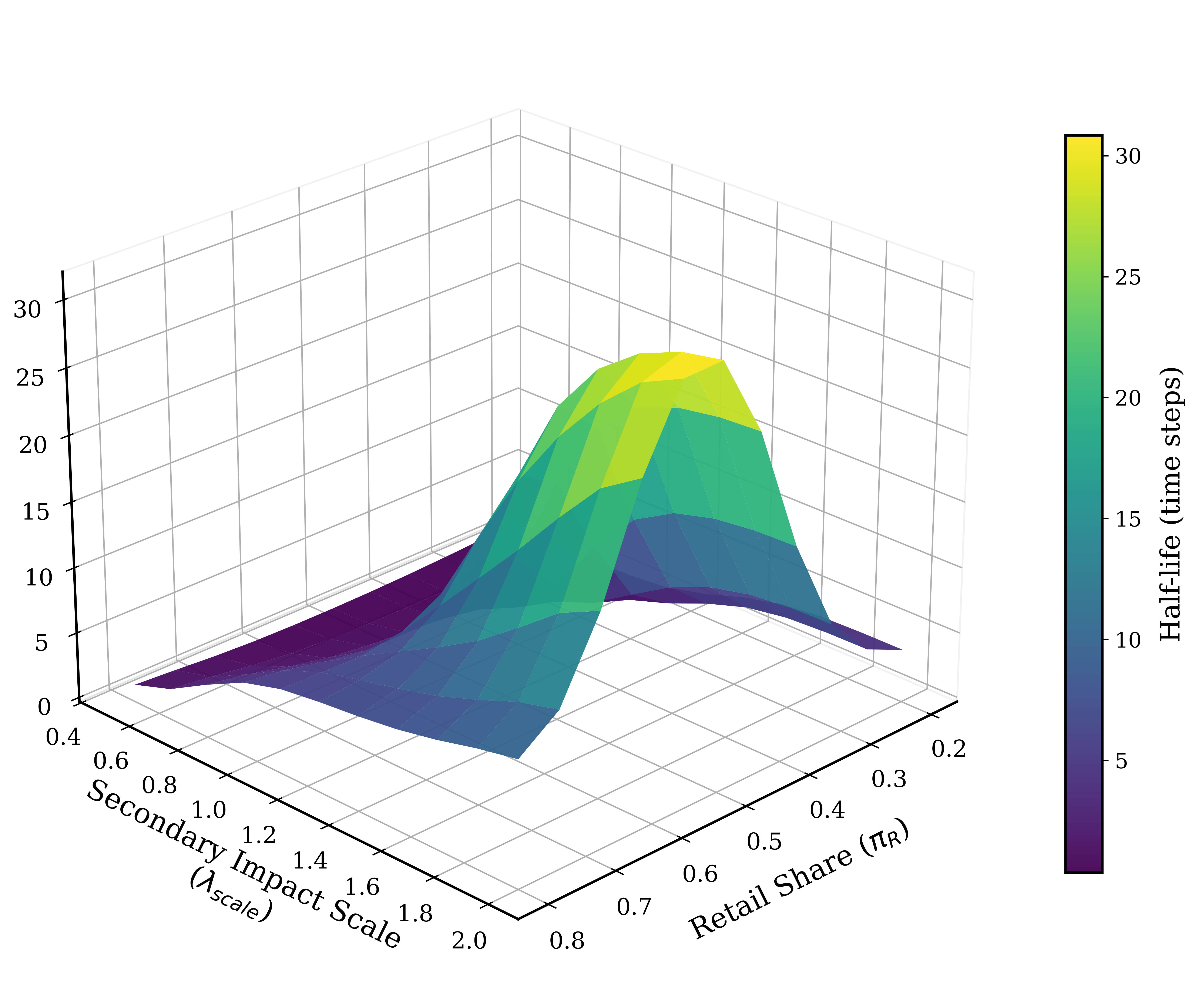}
    \caption{De-peg half-life as a function of retail trader share $\pi_R$ and secondary price impact scale $\lambda_{\text{scale}}$ with fixed primary frictions.}
    \label{fig:secondary_structure_sensitivity}
\end{figure}

\begin{figure*}[!h]
    \centering
    \begin{minipage}[t]{0.32\textwidth}
        \centering
        \includegraphics[width=0.9\linewidth]{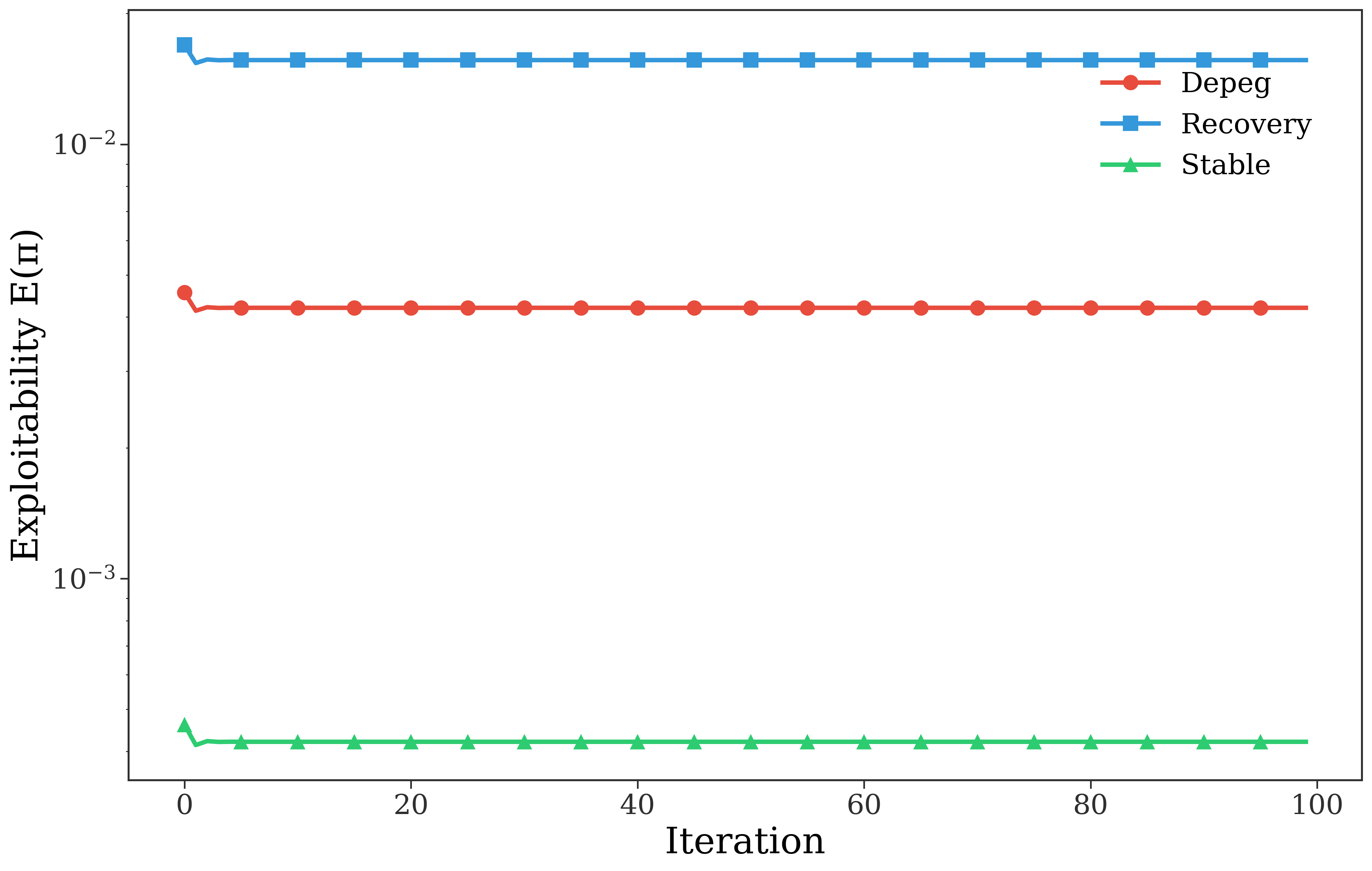}
    \end{minipage}\hfill
    \begin{minipage}[t]{0.32\textwidth}
        \centering
        \includegraphics[width=0.9\linewidth]{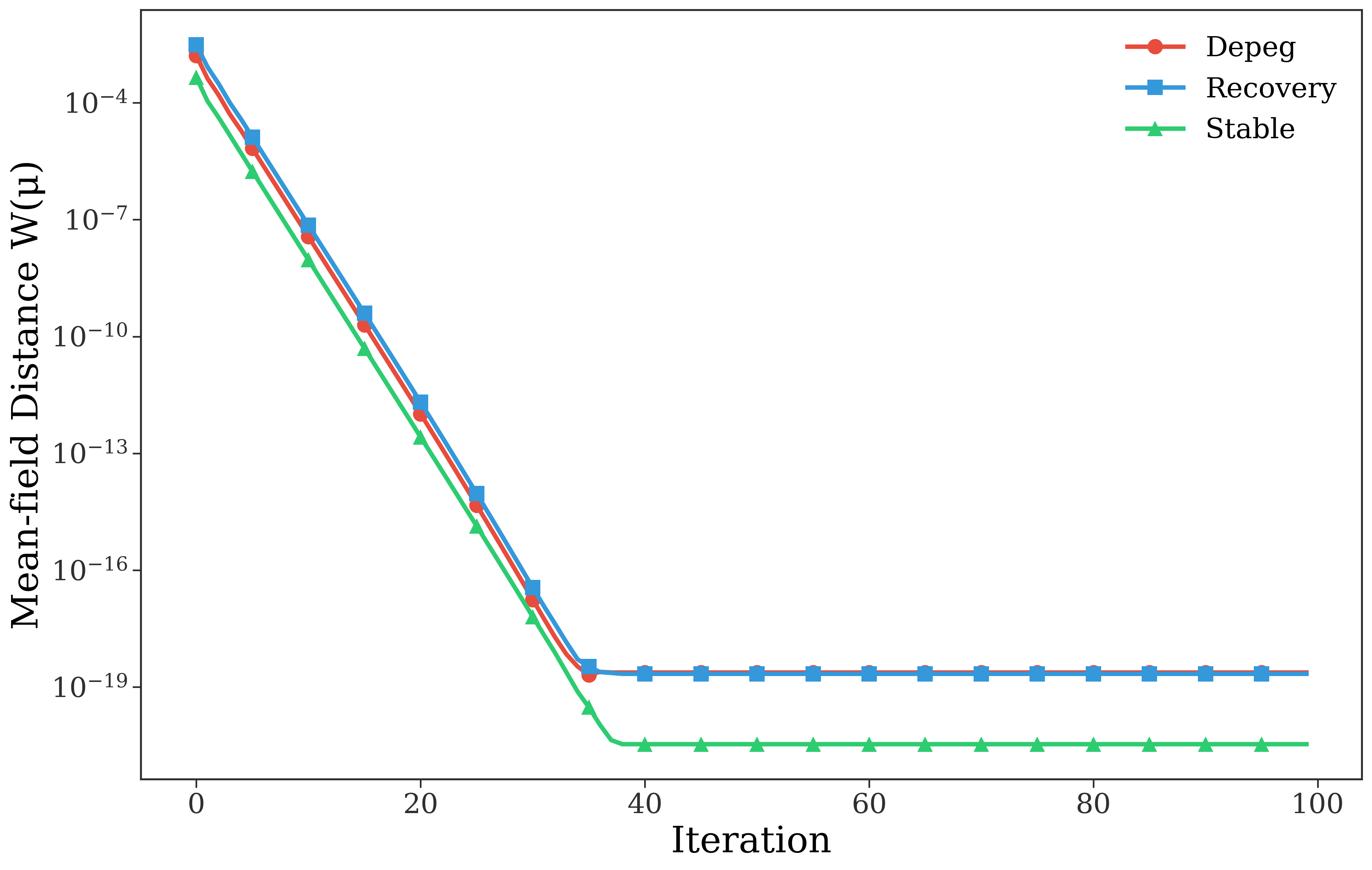}
    \end{minipage}\hfill
    \begin{minipage}[t]{0.32\textwidth}
        \centering
        \includegraphics[width=0.9\linewidth]{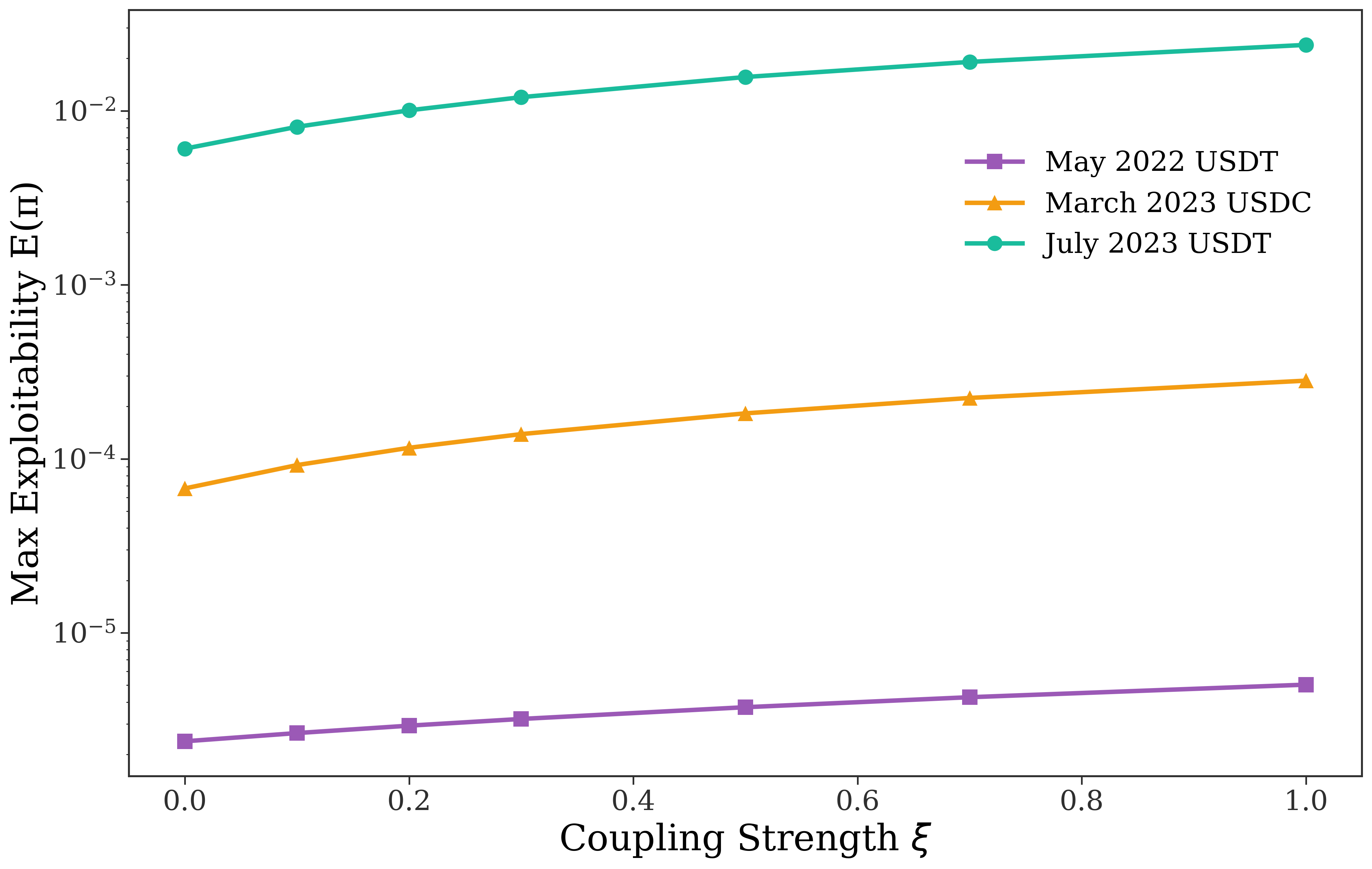}
    \end{minipage}
    \caption{Convergence diagnostics and coupling sensitivity. Left: exploitability $E(\pi)$ across policy-iteration steps. Middle: mean-field distance $W(\mu)$ across steps. Right: maximum exploitability at convergence vs coupling strength $\xi$ across de-peg events.}
    \label{fig:convergence_and_coupling}
\end{figure*}

To evaluate how much explanatory power we gain from the dynamic MFG relative to standard reduced-form approaches, we benchmark it against three time-series models that are widely used to describe mean-reverting price dynamics \cite{tsay2010analysis}. All models are estimated separately for each de-peg episode, using the same calibration windows as those used for the dynamic MFG.
For each de-peg event and each model, we compute two complementary metrics. 
Table~\ref{tab:baseline_summary} reports averages of these metrics across the three calibrated de-peg episodes. Among the models, the dynamic MFG achieves the lowest average RMSE and a lower average half-life error than the reduced form baselines taken collectively, even though the AR(1) and ARMA-GARCH specifications achieve slightly smaller half–life errors in isolation. 

More importantly, only the dynamic MFG explicitly represents interacting retail and arbitrageur populations trading across primary and secondary venues, making it suitable for counterfactual experiments on market structure that are beyond the scope of purely statistical models.
To check that the dynamic MFG does not simply overfit a single calibration window, we also perform a simple out-of-sample evaluation on the March 2023 USDC de-peg. For each of three train/test splits (70/30, 80/20, 90/10), we re-estimate the model on the training segment only and then simulate the price path over the held-out portion of the event. Table~\ref{tab:usdc_oos} reports the resulting root mean squared error and mean absolute error. As the training share increases, in-sample errors decline monotonically, while test RMSE and MAE remain stable in a narrow band around \(7\times 10^{-3}\) and \(6\times 10^{-3}\), respectively. The small variation in test performance across splits suggests that the dynamic MFG captures persistent features of the de-peg dynamics rather than idiosyncratic noise specific to a particular calibration window.

\subsection{Decomposition of Peg-Reverting Flows}

A key objective of our framework is to identify which participants and channels restore the peg after a de-peg. Using the calibrated model, we decompose the total peg-reverting flows into two components (1) Primary Market: net mint/redeem flow, accessible only to arbitrageurs and capturing direct redemptions at par, and (2) Secondary Market: net buying on exchanges by all agents, capturing speculative or exchange-based arbitrage. For each event, we quantify these contributions by integrating aggregate primary redemption flow $\rho_p$ and secondary net buying $\rho_s$ over the simulation horizon.
Figure~\ref{fig:peg_flow_contrib} shows that the dominant restoration channel is event-specific. In the May 2022 USDT episode, stabilization is overwhelmingly driven by primary-market redemptions, consistent with a functional redemption rail and active arbitrage. In the March 2023 USDC episode, the primary channel initially contributes negatively due to impaired minting and redemptions, and recovery reflects a joint role of primary and secondary forces once the primary rails are restored. In the July 2023 USDT venue-level failure, both channels contribute little, consistent with a localized outage that simultaneously blocks primary arbitrage and fragments secondary liquidity.

\subsection{Sensitivity to Primary \& Secondary Market Structure}

Our event calibrations point to primary mint/redeem friction $\kappa_P$ as the main determinant of peg stability, capturing higher redemption fees, tighter access constraints (e.g., KYC or banking limits), and slower settlement on primary rails. We generalize this insight with counterfactual simulations that vary (i) primary versus secondary trading frictions and (ii) market composition under primary-market stress. Figure~\ref{fig:heatmaps_all} shows that $\kappa_P$ is the binding constraint: when the primary channel is inexpensive (roughly $\kappa_P \le 10$), de-peg half-life remains short (about 2-4 model steps) across a wide range of secondary frictions $\kappa_R$ and retail shares $\pi_R$. Once $\kappa_P$ rises past a critical range (about 15-18), half-life increases sharply, consistent with a breakdown of mint/redeem arbitrage, and secondary illiquidity becomes relevant mainly in this impaired-primary region.

Holding primary frictions fixed at a moderately stressed level, we then vary secondary-market structure via retail share $\pi_R$ and a liquidity-depth scale $\lambda_{\text{scale}}$ that rescales secondary price impact. Figure~\ref{fig:secondary_structure_sensitivity} shows a monotone slowdown in recovery as $\lambda_{\text{scale}}$ increases (shallower books), with half-life exceeding 30 steps in the thinnest regimes. The slowdown is most pronounced at intermediate retail shares (approximately $\pi_R \in [0.4,0.7]$), where substantial retail flow combined with limited depth sustains mispricing and raises the cost of arbitrage. Recovery is faster when the population is heavily driven by arbitrageurs, and it moderates again when the market is almost entirely retail, due to reduced aggregate stabilizing capacity.

\subsection{Analysis of Equilibrium Quality and Exploitability}

We next examine the quality of the equilibria produced by the policy iteration solver. For any agent type $i \in \{R, A\}$ and joint stationary policy $\pi = (\pi_R, \pi_A)$, let $J_i(\pi; \mu^\pi)$ denote the infinite horizon discounted cost defined in (1) and (2) when the mean field path $\mu^\pi = \{\mu_t^\pi\}_{t \ge 0}$ is generated by $\pi$ and then held fixed. The cost $J_i(\pi; \mu^\pi)$ is the expected discounted sum of per-period trading PnL and cost terms, measured in dollars per unit notional.
Given a fixed equilibrium environment $(\pi, \mu^\pi)$, the best response $\pi_i^{\mathrm{BR}}$ for type $i$ is obtained by solving the same linear quadratic control problem (\ref{eq:retail_obj}) or (\ref{eq:arb_obj}), but treating the mean field path $\mu^\pi$ and the implied congestion terms as exogenous and keeping the policy of the other population $\pi_{-i}$ fixed. In other words, the deviating population reoptimizes once against the equilibrium environment, while the mean field and the other population do not adjust. We then define the normalized exploitability of $\pi$ for type $i$ as
\begin{equation}
    E_i(\pi)
    =
    \frac{J_i(\pi; \mu^\pi) - J_i(\pi_i^{\mathrm{BR}}, \pi_{-i}; \mu^\pi)}
         {\lvert J_i(\pi; \mu^\pi)\rvert}.
\end{equation}
This dimensionless quantity measures the fractional improvement in the agent's value function that can be obtained by a unilateral deviation, on the same infinite-horizon discounted scale as the original objective. 
Figure~\ref{fig:convergence_and_coupling} summarizes equilibrium quality and convergence. The policy-iteration solver quickly reaches small arbitrageur exploitability $E(\pi)$, with final levels around $10^{-4}$ in Stable, $10^{-3}$ in De-peg, and a few $10^{-2}$ in Recovery, indicating tight $\varepsilon$-Nash behavior under the one-shot deviation test. In parallel, the mean-field path converges rapidly, with the Wasserstein distance $W(\mu)$ between successive trajectories collapsing to a very low stable value, consistent with a strongly contractive fixed-point update across all regimes. Finally, varying the congestion coupling $\xi$ shows that equilibrium quality is robust to stronger mean-field interaction: the maximum converged exploitability remains below a few $10^{-2}$ even at the largest $\xi$ and decays toward $10^{-4}$ as coupling weakens.

\section{Conclusions and Future Work}\label{sec:conclusion}
We develop a dynamic Mean-Field Game framework for fiat-collateralized stablecoins, linking market microstructure, agent incentives, and peg dynamics within a single equilibrium model. Calibrations to three major episodes reproduce key features of observed price paths and half-lives, and allow us to isolate which channel restores the peg under different shocks -- directly addressing the central question of \textit{``who restores the peg?''} in practice. When primary redemption rails remain open, recovery is dominated by direct redemptions. When primary capacity is impaired but not broken, recovery reflects a joint contribution from primary arbitrage and secondary buying. When venue-level infrastructure failures both obstruct redemptions and fragment liquidity, both channels contribute little, and recovery is slow.

A central structural result is a non-linear threshold in primary-market friction. De-peg half-life remains short when primary execution costs stay within a functional range, but rises sharply once primary access becomes sufficiently expensive, beyond which even favorable secondary-market liquidity cannot restore rapid re-pegs. This converts the model into an operational design and risk-management criterion: peg robustness depends not only on reserve quality, but also on the accessibility, speed, fees, and operational resilience of mint and redeem rails, while secondary-market liquidity primarily acts as a complement that limits propagation of localized outages.

An oversimplification is that we work with two representative populations (retail and arbitrageurs) in a tractable LQ equilibrium setting. This abstraction simplifies heterogeneity among liquidity providers and can understate non-linear behavior under extreme stress, including binding constraints and discontinuous primary-rail access. It also treats volatility and liquidity as regime inputs and calibrates primarily to price paths and half-lives, rather than jointly matching flows and order book dynamics. Building on this, future work will incorporate additional agent types and venue-specific constraints, move beyond the LQ specification using learning-based MFG solvers with richer risk objectives and occasionally binding constraints, endogenize liquidity and volatility feedback within the mean field, and validate the model by jointly fitting on-chain flows with high-frequency order book data.

\label{Bibliography}

\bibliographystyle{IEEEtran}
\bibliography{references}

\end{document}